\documentstyle[12pt]{article}

\input{epsf}
\def\be{\begin{equation}}
\def\ee{\end{equation}}
\def\bea{\begin{eqnarray}}
\def\eea{\end{eqnarray}}
 
\newcommand{\lb}{\label}

\renewcommand{\v}{{\bf v}}

\renewcommand{\i}{{\rm i}}
\renewcommand{\u}{{\bf u}}

\renewcommand{\i}{{\rm i}}

\renewcommand{\d}{{\rm d}}

\begin{document}
\baselineskip = 16 pt
\vspace{1.0truein}

\begin{center}
{\Large \bf Odor recognition and segmentation by a model olfactory 
bulb and cortex}

\vspace{.2truein}

{\large Zhaoping Li$^*$ and John Hertz$^\dagger$}

\vspace{.2truein}

{$^*$ Gatsby Computational Neuroscience Unit\\17 Queen Square, 
UCL, London WC1N, 3AR, U.K.  \\zhaoping@gatsby.ucl.ac.uk }

\vspace{.2truein}
  
{$^\dagger$ NORDITA, Blegdamsvej 17,
        2100 Copenhagen {\O}, Denmark  \\hertz@nordita.dk  }

\end{center}

\vspace{1.5truein}
{\centerline{Published in {\it Network: Computation in Neural Systems} {\bf 11} p.83-102. 2000}
\vspace{1.5truein}

\begin{abstract}

We present a model of an olfactory system that performs odor segmentation.
Based on the anatomy and physiology of natural olfactory systems, it consists
of a pair of coupled modules, bulb and cortex.  The bulb encodes the odor 
inputs as oscillating patterns.  The cortex functions as an associative
memory: When the input from the bulb matches a pattern stored in the 
connections between its units, the cortical units resonate in an oscillatory
pattern characteristic of that odor.  Further circuitry transforms this
oscillatory signal to a slowly-varying feedback to the bulb.  This feedback 
implements olfactory segmentation by suppressing the bulbar response to
the pre-existing odor, thereby allowing subsequent odors to be singled
out for recognition.

\end{abstract}

\newpage

\section{Introduction}
 
An olfactory system must solve the problems of odor detection, recognition, 
and segmentation.  Segmentation is necessary because the odor environment 
often contains two or more odor objects.  The system must be able to 
identify these objects separately and signal their presence to higher 
brain areas.  An odor object is defined as an odor entity 
(which, e.g., the smell of a cat, often contains fixed proportions 
of multiple types of odor molecules) that enters the environment 
independently of other odors.  Therefore, two odor objects usually 
do not enter the environment at the same time although they often stay 
together in the environment afterwards.  In cases when different odors
do enter the environment together as a mixture, human subjects have
great difficulty identifying the components\cite{Laing}. In this paper we present a 
model which performs odor segmentation temporally.  First one odor object is 
detected and recognized, then the system adapts to this specific odor 
so a subsequent one can be detected and recognized.

The odor specificity of this adaptation is the key feature of the operation 
of the system.  This specificity can not be achieved with simple single-unit
fatigue mechanisms \cite{HornUsher,HSU} because of the highly distributed 
nature of odor pattern representations in the olfactory system:  fatiguing 
neurons that respond to one odor would strongly reduce their response to 
another one, thereby distorting the pattern evoked by the second odor.  In 
our model a delayed inhibitory feedback signal is directed to the input units 
in such a way as to cancel out the current input, leaving the system free to 
respond to new odors as if the first one were not there.  

Our model is not intended as a faithful representation of any particular
animal olfactory system.  Present anatomical and physiological knowledge
do not permit such detailed modelling.  Rather, our focus is on the 
computations performed by different groups of neurons, based on general
biological findings, which we review briefly here. 

In animals, different odor molecules produce different, distributed 
activity patterns across the neurons of the olfactory nerve, which
provide the input to the olfactory bulb \cite{Shepherd90,Shepherd7990}.
We do not model this part of the processing.  We will simply represent
different odors as different but overlapping input patterns to the bulb. 
They are temporally modulated by the animal's sniff cycle (typically 
2-4 sniffs per second), i.e., active only during and immediately after
inhalation.  
 
The main cell types of the mammalian bulb are the excitatory  mitral cells 
and the inhibitory granule cells.  The mitral cells receive the odor input
and excite the granule cells, which in turn inhibit them.  The outputs
of the bulb are carried to the olfactory cortex by the mitral cell axons. 
In vertebrate animals, odors evoke oscillatory bulbar activity in the
35-90 Hz range, which may be detected by surface EEG electrodes 
\cite{Freeman78,FS}.  Different parts of the bulb have the same dominant 
frequency but different amplitudes and phases \cite{FS,FreemanGrajski},
and this oscillation pattern is odor-specific \cite{FreemanGrajski,Adrian}. 
These oscillations are an intrinsic property of the bulb, persisting after 
central connections to the bulb are cut \cite{FreemanSkarda,GelperinTank}.
(In invertebrates, oscillations exist without odor input but are modulated 
by odors \cite{Delaneyetal}.)  Upon repeated presentation of a conditioned 
odor stimulus, the bulbar oscillations weaken markedly \cite{Bressler}.  
Since olfactory receptor neurons exhibit only limited adaptation 
\cite{Moncrieff, Maetal}, this adaptation must originate either in the
bulb or in cortical structures.

The pyriform or primary olfactory cortex receives bulbar outputs via the 
lateral olfactory tract, which distibutes outputs from each mitral cell
over many cortical locations \cite{Shepherd90}. The signals are conveyed 
to the (excitatory) pyramidal cells of the cortex, both directly and via 
feedforward inhibitory cells in the cortex.  The pyramidal cells send axon 
collaterals to each other and to feedback interneurons which, in turn,
inhibit them.  There is thus excitatory-inhibitory circuitry as in the bulb, 
and oscillatory responses to odors are observed in the cortex, too.  However, 
the cortex differs from the bulb in the much greater spatial range of the 
excitatory connections and in the presence (or at least the greater extent) 
of excitatory-to-excitatory connections.  This anatomical structure has
led a number of workers to model the olfactory cortex as an associative
memory for odors \cite{Haberly85,WB92,AGL,Hasselmo,LilWu,Lilj}.  
Furthermore, the oscillations in the cortex require input from the bulb;
they do not occur spontaneously.  Cortical output, including the feedback 
to the bulb, is from pyramidal cells \cite{Shepherd90}.  Some of the 
feedback is direct, while some of it is via other cortical centers, 
notably the entorhinal cortex.  Most central feedback to the bulb is 
to the granule cells \cite{Shepherd7990}. 
Cooling the cortex, presumably reducing or removing the central feedback,
enhances the bulbar responses \cite{GraySkinner}.

The basic features outlined here constrain our model: we employ coupled
excitatory and inhibitory populations in both bulb and cortex, we
wire the network so that odors evoke oscillations in the bulb, which
drive similar cortical oscillations through excitatory and inhibitory 
connections, and we send the central feedback to reduce the bulbar responses.  

We will neglect many known features of animal olfactory systems, such 
as (to name a few) the patterns of connectivity from receptors to mitral 
cells, the dendrodentritic character of the mitral-granule synapses, and 
the differing spatial range of connectivity in bulb and cortex.   Indeed, 
the model has no geometry: ``location'' and ``distance'' have no meaning 
here.  We retain only the basic elements necessary to illustrate the basic 
operation of the system, in order not to  obscure the functions we focus 
on (detection, recognition, and segmentation). 

We will also hypothesize features of the system, in particular the nature 
of the feedback signal from the cortex to the bulb, for which there is not 
yet experimental evidence (though they are not incompatible with present 
knowledge).  These assumptions will be necessary in order to make an 
explicit model that can be tested computationally.  Some details of its
implementation are neither crucial to the computational function of the
model nor intended as explicit neurophysiological predictions.  However, 
the basic framework of the model and the dynamical properties we find
for it are subject to experimental test. 

In the next section we present the model: its equations of motion and 
how it detects, recognizes, and segments odor inputs.  The following section
demonstrates how it works in simulations.  In the final section we discuss the 
implications of our work, including potential experimental tests for this and
related models and how they can help us understand the functioning of the
olfactory system.

\section{The model}

\begin{figure}[hthththth!]
\begin{center}
\begin{picture}(200, 400)
\put(0, 0)
{\epsfxsize=300pt \epsfbox{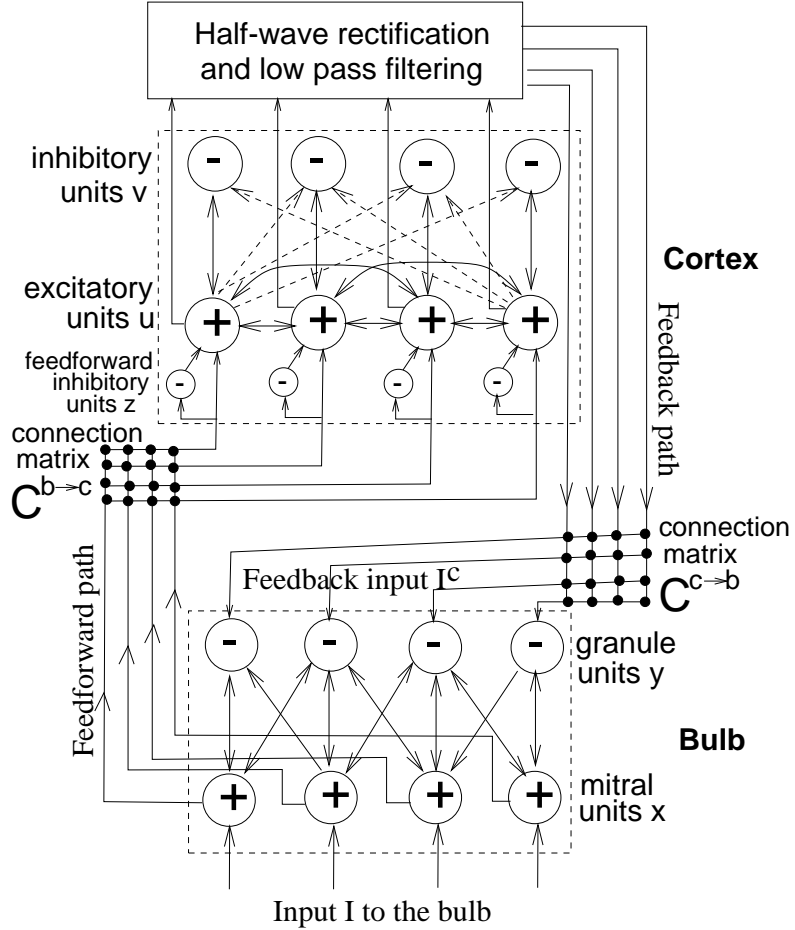}}
\end{picture}
\end{center}
\caption{ \label{fig:olfactorysystem}
\footnotesize{The model.  Odor inputs ${\bf I}$ are fed into the mitral units ($x$) 
in the bulb.  These interact with the inhitory granule units ($y$), both 
locally (vertical connection lines) and nonlocally, via the connection
matrices $\sf H$ and $\sf W$ (diagonal connection lines).  The mitral units 
project their outputs to the cortex via the feedfoward matrix
${\sf C}^{\rm b\rightarrow c}$. The excitatory units in the cortex ($u$)
receive these inputs both directly and indirectly via the feedforward
inhibitory units ($z$). In addition to the local excitatory-inhibitory
connections (vertical lines) between the excitatory ($u$) and the feedback 
inhibitory units ($v$), there are nonlocal connections among the excitatory 
units ($\sf J$, solid lines) and from excitatory to inhibitory units 
($\sf \tilde W$, dotted lines).  The outputs of the excitatory units are 
fed back through a matrix ${\sf C}^{\rm c\rightarrow b}$ to the granule
units in the bulb after rectification and low-pass filtering. (Details of 
the rectification/filtering operation are shown in Fig.\ 
\ref{fig:feedbackroute}.)
}
}
\end{figure}

Our model consists of two modules, a bulb and a cortex, with feedforward
and feedback connections between them.  It is depicted schematically in Fig.\ 
\ref{fig:olfactorysystem}.  The bulb encodes odor inputs as patterns of
oscillation.  These form the input to the cortex, which acts as an associative 
memory for odor objects, recognizing them by resonant oscillation in an
odor-specific pattern when the input from the bulb matches one of its stored 
odor memories.  The odor-specific resonant cortical activity pattern is
transformed to a feedback signal to the bulb, which approximately cancels the 
effect of the odor input that elicited it.  The system is then able to respond 
to a newly arrived odor superposed on the previous one.  In this way it
segments temporally the different odor objects in the environment. 

The model is a rate-model network \cite{WilsonCowan}, in which we associate 
each unit with a local population of cells that share common synaptic input 
(mitral cells for the excitatory units, granule cells for the inhibitory ones). 
The output (activation) of a unit, representing the average firing rate within 
the corresponding population, is modeled as a sigmoidal function of the net 
synaptic input. 

In both the bulb and cortex modules, the units occur in pairs, one unit
excitatory and the other inhibitory.  In the absence of coupling between
different such pairs, they form independent damped local oscillators.  The 
coupling between pairs leads to oscillation patterns across the modules,
with specific amplitudes for the individual local oscillators and specific
phase relations between them.  The odor input makes these oscillatory patterns
different from odor to odor; thus, these patterns form the internal encoding 
of the odors.  The sizes of the local populations corresponding to our formal 
units are different for excitatory and inhibitory units; this difference is 
accounted for in the model by appropriate scaling of the synaptic strengths.  
 
We turn now to the explicit mathematical description of the two modules and
the coupling between them.

\subsection{the bulb}

The bulb model we employ was introduced by Li and Hopfield (1989) 
\cite{LH,Li90}.  For completeness, we review it here.  

The odor input to (mitral) unit $i$ is denoted $I_i$.   (We will
also use a vector notation, in which the entire input pattern is denoted
${\bf I}$.)  Adding to this the synaptic input from granule cells 
within the bulb, we obtain an equation of motion
\be
\dot x_i = - \alpha x_i -\sum_j H^0_{ij}g_y(y_j)+ I_i  \label{eq:1} 
\ee
for the (local population average) membrane potential $x_i$.  Here 
$\alpha^{-1}$ is the membrane time constant, $g_y(\cdot)$ is the (sigmoidal)
activation function  of the granule units, $y_j$ is the membrane potential 
for granule unit $j$, and $H^0_{ij}$ is the inhibitory synaptic strength from 
granule unit $j$ to mitral unit $i$.   All the $H^0_{ij}$ are non-negative; 
the inhibitory nature of the granule cells is represented by the negative 
sign in the second term on the right-hand side.  The signal the bulb sends 
on to the cortex is carried by the mitral unit outputs $g_x(x_i)$ (with
$g_x(.)$ their activation function).  We have not included mitral-mitral 
connections here, because the experimental evidence for them is weak, but 
including them would not change the  properties of the model qualitatively. 

For the inhibitory units, representing local populations of granule cells,
we have, similarly to (\ref{eq:1}),
\be
\dot y_i = - \alpha y_j  +\sum_j W^0_{ij}g_x(x_j) + I^{\rm c}_i,	\lb{eq:1inh}
\ee
with the mitral-to-granule synaptic matrix $W^0_{ij}$.  Here the external 
input $I^{\rm c}_i$ represents the centrifugal input (from the cortex), which 
contains the feedback signal that implements the odor-specific adaptation.  
In describing the response to an initial odor, it can be neglected or taken 
as a constant background input.  

To see how this network produces oscillatory excitation patterns in response
to an odor, start by taking the input ${\bf I}$ to be static.  It determines 
a fixed point $\bar x_i$ and $\bar y_i$ of the equations, i.e., 
$\dot x_i = \dot y_i = 0$ at $\bar x_i$ and $\bar y_i$, which increase with 
odor input ${\bf I}$. Taking the deviation from this fixed point as 
$x_i  - \bar x_i \rightarrow x_i$ and $y_i -\bar y_i \rightarrow y_i$,
linearizing and eliminating the $y_i$ leads to 
\be
\ddot x_i + 2\alpha \dot x_i + \alpha^2 x_i + \sum_j A_{ij}x_j,	\lb{eq:lin}
= 0,
\ee
where the matrix ${\sf A} = {\sf HW}$, with $H_{ij} = H^0_{ij}g'_y(\bar y_j)$ 
and $W_{ij} = W^0_{ij}g'_x(\bar x_j)$.  This equation describes a coupled 
oscillator system, with a coupling matrix $\sf A$.  Denoting the eigenvectors 
and eigenvalues of this matrix by ${\bf X}_k$ and $\lambda_k$, respectively, 
(\ref{eq:lin}) has solutions ${\bf x} = \sum_k c_k {\bf X}_k {\rm exp}[-\alpha 
t \pm {\rm i}(\sqrt {\lambda_k} t + \phi_k)]$, with $c_k$ and $\phi_k$ the 
amplitude and phase of the $k^{th}$ mode.  If $\sf A$ is not symmetric (the
general case), $\lambda_k$ is complex, and the mode has oscillation 
frequencies $\omega _k \equiv {\rm Re} (\sqrt {\lambda_k })$.  The 
amplitude for mode $k$ will grow exponentially (in this linearized theory) 
if $\pm {\rm Im} ({\sqrt \lambda_k })> \alpha$.  Its growth will be limited
by nonlinearities, and it will reach a steady-state saturation value.  In 
this spontaneously oscillating state, the fastest-growing mode, call it the 
$1^{st}$ mode, will dominate the output. The whole bulb will oscillate with 
a single frequency $\omega _1$ (plus its higher harmonics), and the 
oscillation amplitudes and phases may be approximated by the complex 
vector ${\bf X}_1$.  Thus, the olfactory bulb encodes the olfactory 
input via the following steps: (1) the odor input $\bf I$ determines the 
fixed point $({\bf \bar x}, {\bf \bar y})$, which in turn (2) determines 
the matrix ${\sf A}$, which then (3) determines whether the bulb will 
give spontanous oscillatory outputs and, if it does, the oscillation 
amplitude and phase pattern ${\bf X}_1$ and its frequency $\omega_1$.

Strictly speaking, this description only applies to very small oscillations.
For larger amplitudes, nonlinearities make the problem in general 
intractable.  However, we will suppose that the present analysis gives a
decent qualitative guide to the dynamics, checking this assumption later
with simulations of the network.

In this model, oscillations arise strictly as a consequence of the
asymmetry of the matrix $\sf A$.  The model could be generalized to add
intrinsic single-unit oscillatory properties, and these might
enhance the 
network oscillations.  However, a model with symmetric $\sf A$ and intrinsic 
oscillatory properties only at the single-unit level can not support 
oscillation patterns in which the phase varies across the units in the 
network.  We will return to this point in the Discussion section.

A word about timescales: The odor input varies on the timescale of a 
sniff: 300-500 ms.  The oscillations are in the 40 Hz range, so 
the input ${\bf I}$ hardly changes at all over a few oscillation periods ($\sim 25$ ms).
We may therefore treat periods of several oscillations as if the input were
static within them, and do the above analysis separately for each such
period (adiabatic approximation). 

With inhalation, the increasing input ${\bf I}$ pushes the fixed point 
membrane potentials 
$\bar x_i$ from their initial values (where the activation function $g(x)$ 
has low gain) through a range of increasing gains, thereby increasing the size 
of some of the elements of the matrix $\sf A$ (recall the definition of
$\sf A$ above).  This increases the magnitude of both the real and imaginary 
parts of the eigenvalues $\lambda_k$, until the threshold where $|{\rm Im} 
({\sqrt \lambda_k })| = - \alpha$, where oscillations appear.  These oscillations increase 
in amplitude as the input increases further, until the animal stops inhaling 
and the input ${\bf I}$ decreases.  Then the oscillations shrink and disappear as the system returns
toward its resting state.   
This rise and fall of oscillations within each sniff cycle give the bulb
outputs both a slowly-varying component (2-4 Hz) and a high frequency 
(25-60 Hz) one, as observed experimentally \cite{FS}.

It is not known how the synaptic connections represented in the model by 
the matrices ${\sf H}^0$ and ${\sf W}^0$ develop in the real olfactory bulb, 
and we do not attempt to model this process here.  It is possible that the 
real bulb acts, to some degree, as an associative memory as a result of this 
learning.  However, our conclusions will not depend on this.  Similarly, 
our analysis does not depend on details of the synaptic matrices, such as 
their range and degree of connectivity.  We require only that the connections 
lead to distinct oscillation patterns for different odors, with dissimilar 
patterns evoked by dissimilar odors.

\subsection{the cortex}

Our cortical module is structurally similar to that of the
bulb. However, there are the following significant differences:
(1) The cortex receives an oscillatory input from the bulb, while the
bulb receives non-oscillatory (at the time scale of the cortical
oscillation) input; (2) The cortex has excitatory-to-excitatory connections, 
while our bulb module does not.

We  focus on the local excitatory (pyramidal) and feedback inhibitory 
interneuron populations. The units that represent them obey equations of 
motion similar to those for the mitral and granule units of the bulb:
\bea
\dot u_i &=& -  \alpha u_i  -\beta^0 g_v(v_i)
+ \sum_{j} J^0_{ij} g_u(u_j)  
-\sum_{j} \tilde H^0_{ij} g_v(v_j) + I^{\rm b}_i, 
					\lb{eq:ueqn}	  \\
\dot v_i &=& - \alpha v_i  +\gamma^0 g_u(u_i)
+\sum_{j} \tilde W^0_{ij} g_u(u_j).		\lb{eq:veqn}	
\eea
Here $u_i$ represent the the average membrane potentials of the local 
excitatory populations and $v_i$ those of the inhibitory populations. 
The synaptic matrix ${\sf J}^0$ is excitatory-to-excitatory connections, 
${\sf \tilde H}^0$ is inhibitory-to-excitatory connections, and 
${\sf \tilde W}^0$ is excitatory-to-inhibitory connections.  For later
convenience, we have written the local terms (the effect of $v_i$ on $u_i$ 
and vice versa) explicitly, so ${\sf \tilde H}^0$ and ${\sf \tilde W}^0$ 
have no diagonal elements.  We also assume $J^0_{ii} = 0$.  $I^{\rm b}_i$ 
are the net inputs from the bulb, both directly and indirectly via the
feedforward inhibitory units (see later for the description of this
pathway).  Like the bulb activity itself, these contain in general both
a slow part $I^{{\rm b}0}_i$, varying with the sniff cycle, and an 
oscillating ($\gamma$-band) part $\delta I^{\rm b}_i$, i.e., 
$I^b_i \equiv I^{{\rm b}0}_i + \delta I^{\rm b}_i$. 

We can carry out the same analysis as in the bulb, taking the fixed point 
as $({\bf \bar u}, {\bf \bar v})$, which are determined by
${\bf I}^{{\rm b}0}$, i.e., $\dot {\bf u} = \dot {\bf v} =0$ at 
$({\bf \bar u}, {\bf \bar v})$ when $I^b_i = I^{b0}_i$ with 
$\delta I^b_i =0$.  Taking  ${\bf u} \rightarrow {\bf u} -{\bf \bar u}$, 
${\bf v} \rightarrow {\bf v} -{\bf \bar v}$, linearizing and eliminating 
the $v_i$, we obtain
\bea
\ddot u_i &+& \sum_j [ 2\alpha \delta_{ij} - J_{ij}] \dot u_j \nonumber \\
          &+& \sum_j [ (\alpha^2 + \beta_i \gamma_i ) \delta _{ij}
- \alpha J_{ij} + \gamma_i \tilde H_{ij} + \beta_i \tilde W_{ij}
+\sum_k  \tilde H_{ik} \tilde W_{kj} ] u_j  
= (\partial_t +\alpha)\delta I_i^{\rm b}.			\lb{eq:cortosc}
\eea
Here $\beta_i = \beta^0 g'_v(\bar v_i)$, $\gamma_i = \gamma^0 g'_u
(\bar u_i)$, $J_{ij} = J^0_{ij}g'_u(\bar u_j)$, $\tilde H_{ij} = \tilde 
H^0_{ij}g'_v(\bar v_j)$, and $\tilde W_{ij} = \tilde W^0_{ij}g'_u
(\bar u_j)$.  Thus this is a system of driven oscillators coupled by 
connections ${\sf J}$, ${\sf \tilde H}$, and ${\sf \tilde W}$ and driven 
by an external oscillatory signal  
$\delta \dot{\bf I}^{\rm b} + \alpha \delta {\bf I}^{\rm b}$, 
which is proportional to $\delta {\bf I}^{\rm b}$ for a purely 
sinusoidal oscillation.  A single dissipative oscillator driven by an 
oscillatory force will resonate to it if the frequency of the driving 
force matches the intrinsic frequency of the oscillator. A system of 
coupled oscillators has its intrinsic oscillation patterns --- the normal 
modes determined by the coupling.  Analogously, it will also resonate to 
the input when the driving force, a complex vector proportional to 
$\delta {\bf I}^{\rm b}$, matches one of the intrinsic modes, also 
a complex vector, in frequency and in its pattern of oscillation amplitudes 
and phases.

It is apparent from Eq.\ (\ref{eq:cortosc}) that the matrices $\sf \tilde H$ 
and $\sf \tilde W$ play the same roles.  Therefore, for simplicity, we will
drop the inhibitory-to-excitatory couplings $\sf \tilde H$ from now on,
thinking of the fact that the real anatomical long-range connections appear 
to come predominantly from excitatory cells.

\subsubsection*{Odor selectivity and sensitivity}

In our model, the olfactory cortex functions as an associative memory, as
described and modeled by a number of authors 
\cite{Haberly85,WB92,AGL,Hasselmo,LilWu,Lilj}.
It is similar to a Hopfield model, but instead of stationary patterns it
stores oscillating patterns which vary in phase as well as magnitude across
the units of the network. 
The memory pattern for the $\mu^{th}$ odor is described by a complex 
vector $\xi^\mu$, whose component $\xi^\mu_i$ describes both the relative 
amplitude and phase of the oscillation in the $i^{th}$ unit. The cortex
stores the memories about the odours in the synaptic weights 
${\sf J^0}$ and ${\sf \tilde W^0}$, or, effectively, the coupling 
between oscillators.
It then recognizes the 
input odors, as coded by the oscillating input patterns 
$\delta {\bf I}^{\rm b}$ (which are linearly related to the bulbar 
oscillatory outputs), by resonating to them, giving high-amplitude 
oscillatory responses itself. If, however, the input
$\delta {\bf I}^{\rm b}$ does not match one of the stored odor 
patterns ${\bf \xi}^\mu$ closely enough, the cortex will fail to respond 
appreciably.  

In the present model the memory pattern $\xi_i^\mu$ for odors 
$\mu = 1, 2, ...$ are designed into the synaptic connections ${\sf J}$ 
and $\sf \tilde W$.   Let $\omega $ be the oscillation frequency, 
$\delta I_i^{\rm b} \propto e^{-{\rm i}\omega t}$.   Once the oscillation 
reaches a steady amplitude $u_i \propto  e^{-{\rm i}\omega t}$,  we have 
$\dot{u}_i = -{\rm i}\omega u_i$, $\ddot u_i = -{\rm i}\omega \dot u_i$, 
so we get
\begin{equation}
\dot u_i  
= [  -  2\alpha   - {{\rm i}\over \omega }  (\beta_i \gamma_i + \alpha ^2)] 
	u_i +\sum_j [ J_{ij}  - {{\rm i}\over \omega } (  
	\beta_i \tilde W_{ij} - \alpha  J_{ij} ) ] u_j
      +  {{\rm i}\over \omega }
      (-\i \omega + \alpha) \delta I^{\rm b}_i. 	\label{eq:oscJW}
\end{equation}
The second term [...] on the right hand side gives an effective coupling 
between the oscillators.  From now on in this analysis we will make the
approximation that the different local oscillators have the same natural 
frequencies, i.e.\ $\beta_i \gamma_i$ is independent of $i$.  
Assuming further that the oscillation frequencies for different odors 
are nearly the same, the odor patterns can then be stored in the matrices 
in a generalized Hebb-Hopfield fashion as
\begin{equation}
        M_{ij}\equiv [ J_{ij}  - {{\rm i}\over \omega } ( 
	\beta \tilde W_{ij} - \alpha  J_{ij})]
        = J \sum_\mu \xi^\mu_i \xi^{\mu*}_j ,		\label{eq:MHebb}
\end{equation}
or, with $\xi_i^{\mu}$ expressed in terms of amplitudes and phases
as $|\xi_i^{\mu}| \exp (-{\rm i}\phi_i^{\mu})$,
\begin{eqnarray}
         J_{ij} &=& J \sum_{\mu} |\xi_i^{\mu}| |\xi_j^{\mu}|
		\cos (\phi_i^{\mu} -\phi_j^{\mu}) 	\label{eq:HebbJ} \\
 \beta \tilde W_{ij}
		&=& J \sum_{\mu} |\xi_i^{\mu}| |\xi_j^{\mu}|
	[\omega \sin (\phi_i^{\mu} -\phi_j^{\mu})
	+\alpha \cos (\phi_i^{\mu} -\phi_j^{\mu})].	\label{eq:HebbHW}
\end{eqnarray}
Note that here both kinds of connections, $\sf J$ (excitatory-to-excitatory) 
and $\sf \tilde W$ (excitatory-to-inhibitory), are used to store the 
amplitude and phase patterns of the oscillation.  $\sf J$ is
symmetric, while $\sf \tilde W$ is not.  

These connections can be obtained by an online algorithm, a simplified
version of the full Hebbian learning treated by Liljenstr\"om and Wu
\cite{LilWu,Lilj}. Suppose the cortex has effective oscillatory input
$\delta {\bf I}^b = {\bf \xi}^\mu e^{-\i\omega t} + {\bf \xi } ^{\mu *}e^{\i\omega t}$ during
learning of the $\mu^{th}$ pattern. Here we make explicit the real nature 
of the signals. Suppose also that the $\sf J$ and $\sf \tilde W$ connections
inactive, consistent with the picture proposed by Wilson, Bower and Hasselmo
\cite{WB92,Hasselmo}, who suggested that learning occurs when the long-range 
intracortical connections are weakened by neuromodulatory effects. Then the 
linearized (4) and (5) are simply
\bea
\dot u_i + \alpha u_i &=& -\beta v_i
+ \xi_i^\mu e^{-\i \omega t}  + \xi_i ^{\mu *}e^{\i\omega t}\nonumber \\
\dot v_i +\alpha v_i &=& \gamma u_i,
\eea
with solution
\bea
u_i(t) &=& \frac{-\i \omega + \alpha}
{-\omega^2 +\alpha^2 +\beta \gamma -2\i \alpha \omega}
\xi_i^{\mu} e^{-\i \omega t}      + {\rm c.c.} \nonumber \\
v_i(t) &=& \frac{\gamma}
{-\omega^2 +\alpha^2 +\beta \gamma -2\i \alpha \omega}
\xi_i^{\mu} e^{-\i \omega t}        +   {\rm c.c.}    \label{eq:solns}
\eea
where {\rm c.c.} denotes complex conjugate. In other words, the cortical activities are clamped by the inputs.

For Hebbian learning, $\dot J_{ij} \! \propto \! u_i(t) u_j(t)$, and, after 
time averaging,
$\delta J_{ij} \! \propto \! \int _{0}^{2\pi /\omega } u_i(t) u_j(t) \d t$,
leading to
\bea
\delta J_{ij} &\propto& \frac{\omega^2 + \alpha^2}
{|-\omega^2 +\alpha^2 +\beta \gamma -2\i \alpha \omega|^2}
(\xi_i^{\mu} \xi_j^{\mu *} + \xi_i^{\mu *} \xi_j^{\mu} ) \nonumber \\
&= &2 \frac{\omega^2 + \alpha^2}
{|-\omega^2 +\alpha^2 +\beta \gamma -2\i \alpha \omega|^2} |\xi^\mu_i| |\xi^\mu_j|
\cos (\phi^\mu_i -\phi^\mu_j).
\eea
Similarly, 
$\delta W_{ij} \propto \int _{0}^{2\pi /\omega } v_i(t) u_j(t) \d t$ 
leading to
\bea
&\delta W_{ij}& \propto  \frac{\gamma}
{|-\omega^2 +\alpha^2 +\beta \gamma -2\i \alpha \omega|^2}
[(\i \omega + \alpha) \xi_i^{\mu} \xi_j^{\mu *}
+(-\i \omega + \alpha) \xi_i^{\mu *} \xi_j^{\mu} ] \nonumber \\
&=&  \frac{2\gamma}
{|-\omega^2 +\alpha^2 +\beta \gamma -2\i \alpha \omega|^2}
[ \omega |\xi^\mu_i| |\xi^\mu_j| \sin  (\phi^\mu_i -\phi^\mu_j)
+ \alpha  |\xi^\mu_i| |\xi^\mu_j|
\cos (\phi^\mu_i -\phi^\mu_j)].
\eea
Then, if the relative learning rates for $\sf J$ and $\sf \tilde W$ are
tuned appropriately, we simply recover the formulae (\ref{eq:HebbJ}) and 
(\ref{eq:HebbHW}).  In actual online learning, we can use high-pass 
versions of $\bf u$ and $\bf v$ to learn $\sf J$ and $ \sf \tilde W$ 
to remove the baseline value, i.e., the operation point $\bar \u$ and 
$\bar \v$, which does not contain odor information.

To see the selective resonance explicitly, suppose that different patterns 
${\bf \xi}^\mu$ are orthogonal to each other.  Let us denote the overlap 
$(1/N) \sum_i \delta I^{\rm b}_i \xi_i^{\lambda *}$ of the input  
$\delta I^{\rm b}_i$ with the stored pattern $\xi_i^{\lambda }$ by
$\delta I^{\lambda}$.  Then, multiplying (\ref{eq:oscJW}) by 
$\xi_i^{\lambda *}$ and summing on $i$, we find that at steady oscillatory
state, the response $u^{\lambda} \equiv (1/N) \sum_i u_i \xi_i^{\lambda *}$ 
to pattern ${\bf \xi}^\lambda$  obeys 
\begin{equation}
\dot u^{\lambda} = -(2\alpha -J) u^{\lambda}  
   - {{\rm i}\over \omega }  (\beta \gamma + \alpha ^2) u^{\lambda}
      +  {{\rm i}\over \omega }
      (-\i \omega + \alpha) \delta I^{\lambda}          
\end{equation}
This is like an oscillator with oscillation frequency 
$(\beta \gamma + \alpha ^2)/\omega$ and an effective oscillation decay rate
$2\alpha -J$. It resonates to external oscillatory input of 
frequency $\omega \approx 
\sqrt {\beta \gamma + \alpha ^2}$  with a steady state amplitude
\begin{equation}
u^\lambda = {{ (-\i \omega +\alpha) \delta I^\lambda } \over
	{ \beta \gamma + \alpha ^2 - \omega^2    
	- {\rm i}\omega (2\alpha -J)}} \approx 
	{{ (1+i\alpha /\omega ) \delta I^\lambda } \over {2\alpha -J}}
\label {eq:ulambda1},
\end{equation}
However, for an input $\delta I^{\rm b}_i$ orthogonal to all the stored patterns,
$\delta I^\lambda = 0$ for all $\lambda $, and the resonance will be washed out
when $J<2\alpha $.  For $J>2\alpha$, the network will support spontaneous 
oscillations analogous to those in the bulb, but not as observed in the cortex. 
The effect of the long-range couplings, through the parameter $J$, is to reduce the 
damping in the circuit from $2\alpha $ to $2\alpha -J$ when the input matches 
a stored pattern, thereby sharpening the resonance as
$J \rightarrow 2\alpha$ while we keep $J<2\alpha $.
On the other hand, the resonant driving frequency depends only on the 
single-oscillator parameters $\alpha$, $\beta$ and $\gamma$.  

This oscillatory associative memory enjoys the usual properties that 
characterize Hopfield networks \cite{Hopfield}, including rapid convergence 
(a few oscillation cycles if the presented pattern has reasonable overlap 
with a stored one), robustness with respect to noise and corrupted input, 
and a storage capacity of the order of $N$ random patterns, where $N$ is
the network size.

\subsection{Coupling between bulb and cortex}

The model has both feedforward (bulb-cortex) and feedback (cortex-bulb)
connections.  The former transmit the bulbar encoding of the input odors
to the cortex for recognition, while the latter permit segmentation by
producing adaptation to recognized odor objects.

\subsubsection*{bulb to cortex}

As mentioned in the Introduction, in the real cortex, the excitatory cells
receive input from the bulb both directly from the fibers of the lateral
olfactory tract and in a slower pathway via feedforward inhibitory 
interneurons in the  cortex.  We model this in the following way.  The
synapses from local bulb populations $j$ to local cortical populations $i$ 
are specified by a matrix $C^{\rm b\rightarrow c}_{ij}$.  The values of these 
connections are not important in the model, and very little is know about 
them, so we will take them to be random.   The resulting signals are 
then fed to the excitatory cells, both directly and, with the opposite
sign, through a parallel low-pass filter, representing the effect of 
the feedforward inhibitory cells; see Fig.\ \ref{fig:olfactorysystem}.
Details are given in the appendix.

The combination of the direct excitatory and low-pass filtered inhibitory
signals makes the feedforward pathway act as a high-pass filter, partially
cancelling the slow part ${\bf I}^{{\rm b}0}$ of the bulb output from the 
cortical input.   Consequently, the net input to the cortical excitatory
units is dominated by the oscillatory component of the bulb activity, which 
encodes information about the odor input.  (We do not know how well such a 
cancellation is actually achieved in real olfactory systems, but this 
could be tested experimentally.)

\subsubsection*{cortex to bulb}
 
The odor-specific adaptation that forms the basis for odor segmentation
in our model is implemented using a feedback signal from the cortex to
the granule units of the bulb.  We do not know how such a signal is
generated in animals, or even whether it is, although anatomically such
a pathway exists.  If the signal does exist, it likely also involves 
areas such as entorhinal cortex, which contributes to the centrifugal input 
to the bulb.  These areas lie outside the scope of the present model, so 
we will simply construct a suitable signal and explore the consequences.

In exploratory computations, we have found that this form of feedback
control only works if the signal is slowly varying in time (on the order 
of the sniff cycle time or slower).  Merely feeding back the oscillating 
cortical activities does not appear to permit any kind of robust 
stimulus-specific adaptation.  

Thus, we generate the feedback signal in the following {\em ad hoc}
fashion:  First each excitatory cortical output $g_u(u_i)$ is run 
through a threshold-linear element to remove its non-oscillatory part,
which carries no odor information.  Then the output of this element is
run through a low-pass filter.  The time constants of this filter are
on the order of the sniff cycle or longer.  The net result is a
signal pattern which takes a sniff-cycle time or so to grow to full
strength.  The signal component from excitatory unit $i$ will be proportional
to the amplitude of the oscillation of that unit, so this signal will
contain information about the odor that evoked the cortical oscillation
pattern.  The explicit form of the equations used to generate the feedback 
signal in the simulations is given in the Appendix.

Since we rectify and low-pass only the excitatory cortical outputs $g_u(u_i)$, 
the feedback signal includes only the odor information coded in the amplitude 
but not in the phase pattern of the cortical oscillation.  Phase information 
could be included by (for example) feeding the difference signals 
$g_u(u_i)-g_u(u_j)$ through the rectification and low-pass processes.  However, 
we have not explored such mechanisms in this work. 

The granule units in the bulb respond to the feedback signals by changing 
their activities proportional to it.  These changes are then transmitted to
the mitral cells by the synaptic matrix $\sf H$.  As shown by Li 
\cite{Li90}, a feedback signal  
\begin{equation}
 {\bf F} \propto {\sf H}^{-1} {\bf I}, 
\end{equation}
will, when transmitted onward to the mitral units, cancel the odor inputs to 
the bulb (in linear approximation).

In our model we want to make this cancellation work for all the odor patterns
stored in the cortex.  Denoting by $G_j^\mu$ the rectified and low-passed 
cortical output when the system is stimulated by odor pattern $I_k^{\mu}$, 
this can be achieved by a Hebbian feedback connection matrix 
${\sf C}^{\rm c\rightarrow b}$ that maps ${\bf G}^\mu$ to feedback signal 
${\bf F}^\mu$
for each odor $\mu$ in a single layer network: 
\begin{equation}
C^{\rm c\rightarrow b}_{ij} \propto
\sum_{\mu} F^\mu _i G^\mu _j = 
\sum_{k}H^{-1}_{ik} \sum_{\mu}I^{\mu}_k G_j^{\mu}. 	\label{eq:fb}
\end{equation}

\section{Simulations}

We have simulated a network with bulb and cortical modules each 
consisting of 50 excitatory and 50 inhibitory units.   They were
coupled as described in Sect.\ 2.3 and the Appendix.
The coupled differential equations were integrated using a
fourth-order Runge-Kutta routine from Numerical Recipes \cite{NR}.

We used three random odor input patterns $I_i^{\mu}$.  Their elements
were drawn independently for each $i$ and $\mu$ from a uniform distribution 
on (0,1].  The elements of the granule-to-mitral synaptic matrix $\sf H$ 
were taken to have the form $H_{ij} = {\rm const.}\cdot \delta_{ij}$.
We designed the mitral-to-granule matrix $\sf W$ so as to 
make the bulb oscillate in response to the three input patterns, taking
$W_{ij} \propto {\rm Im}\, \sum_{\mu=1}^3 \zeta_j^\mu \zeta_j^{\mu *}$.  
Here the $\zeta_i^\mu$ are complex, with amplitudes resembling the input
odor patterns $I_i^{\mu}$ and with random phases. Since  $\sf W$ should have 
non-negative elements, we simply zeroed out the negative $W_{ij}$ in the 
construction.\footnote{In the bulb model of Li and Hopfield\cite{LH,Li90}, 
the idea was that extensive asymmetric random synapses would, for a large 
network, automatically generate a distributed encoding of odors in the
amplitudes and phases of oscillation patterns in the network.  Here we will 
be more concerned with how these patterns are processed by the cortex, so,
for convenience, we have engineered particular amplitude patterns in through 
the bulbar {\sf W} matrix in this fashion.  However, the particular forms 
used for $\sf H$ and $\sf W$ are not important for the problem 
that we are studying here, as long as ${\sf A} \equiv {\sf HW}$ is 
sufficiently asymmetric.} 
This dilution did not affect the bulb oscillations qualitatively. 
Other parameters were set as in \cite{LH}, so the evoked oscillations 
were in the 40-Hz range.  

The cortical design followed Sect.\ 2.2.   The local couplings $\beta^0$ 
and $\gamma^0$ were chosen so that the cortical oscillation frequency 
roughly matched the bulbar one, i.e., 
$\beta^0\gamma^0+\alpha^2 \approx \bar \omega^2$ (see equation 
(\ref {eq:ulambda1}), where $\bar \omega$ is the average oscillation 
frequency in the bulbar outputs.  The inhibitory units had the 
sigmoidal activation function used in the model of the bulb \cite{LH}.
In some of our simulations, the activation function of the excitatory units 
also had this form.  In obtaining the results presented here, however, we 
used a piecewise linear activation function with gains of 1 and 2, 
respectively, in the low- and high-input regions above threshold.  This
choice was made only for convenience in analyzing the nonlinear dynamics 
and is not essential for the function of the network.

The cortical synaptic matrices  $\sf J$ and $\sf \tilde W$ were designed
to store oscillation patterns for two of the three odor input patterns, in
the following way.  For each of the two odors, we stimulated the bulb with 
its input pattern $I_i^{\mu}$ and fed the resulting oscillatory bulb output 
through the bulb-to-cortex matrix $C^{\rm b\rightarrow c}_{ij}$ and the
subsequent high-pass filtering operation to the cortex, with the intracortical
connections ${\sf J}^0$ and ${\sf \tilde W}^0$ set to zero.  The resulting
oscillation patterns in the cortical units for the two odors were then
used as $\xi^\mu$ in constructing $\sf J$ and $\sf \tilde W$.   

We modified the Hebb rule (eq.\ (\ref{eq:MHebb}) or eqs.\ (\ref{eq:HebbJ}) 
and (\ref{eq:HebbHW})) slightly, using, instead, a pseudoinverse formula
\begin{equation}
        M_{ij} 
        = J \sum_\mu \xi^\mu_i \eta^{\mu*}_j ,		\label{eq:Mproj}
\end{equation}  
where $\sum_i \eta^{\mu*}_i \xi^{\nu}_i = N\delta_{\mu \nu}$.   
This was done only to reduce finite-size effects due to mutual overlaps  
(of order $\sqrt{N}$) between patterns, and would be inessential in 
sufficiently large networks. 

As explained in section 2.3 and the Appendix, the slowly-varying feedback 
signal used for the odor-specific adaptation was generated by a 
threshold-linear rectification, followed by a pair of simple linear filters.  
The time constants of these (3 and 0.3 sec respectively) would made it take 
10-12 256-ms sniff cycles to generate a full strength feedback signal if the
cortical signal were held constant.  Similarly, the adaptation takes just as
long to wear off after the stimulus is removed.

Like the intracortical ${\sf M}$ matrix, the cortex-to-bulb matrix 
${\sf C}^{\rm c\rightarrow b}$ was modified using the projection-rule 
algorithm to eliminate finite-size overlap effects between the cortical
oscillation patterns of different odors.  Thus, in the formula (\ref{eq:fb}), 
we replaced the rectified and low-pass-filtered cortical patterns 
$G_j^{\mu}$ by $\tilde G_j^{\mu}$, where $\tilde{\bf G}^{\mu}$ are vectors 
such that $\tilde{\bf G}^{\mu} \cdot{\bf G}^{\nu} = N\delta^{\mu \nu}$. 

Fig.\ \ref{fig:3odors} shows the bulbar and cortical oscillatory 
response patterns evoked on 5 of the  50 mitral or cortical excitatory
units by three odors: A, B, and C.   Only odors A and B are stored 
in the cortical memory in the $\sf J$ and $\sf \tilde W$ matrices. 
Different amplitude response patterns to different odors are apparent.
The cortex resonates appreciably to only odors A or B, but not to C, 
demonstrating the selectivity of the cortical response.

\begin{figure}[hthththt!]
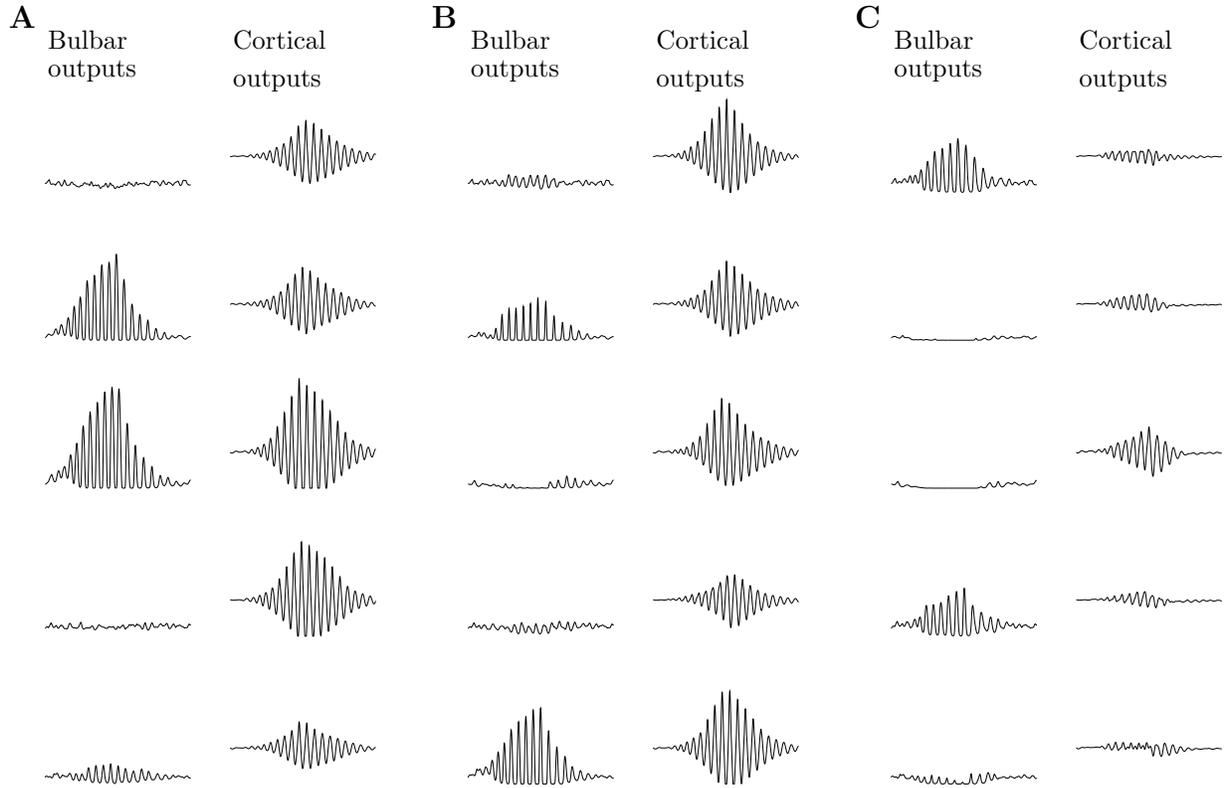

\begin{center}
\setlength{\unitlength}{0.8pt}
\begin{picture}(600, 360)
\put(400,360) {\bf C}
\put(400,0)
{
\setlength{\unitlength}{0.7pt}
\begin{picture}(200, 400)
\put(10, 400) {\footnotesize {Bulbar}}
\put(10, 385) {\footnotesize {outputs}}
\put(0, 320)
{\epsfxsize=63pt \epsfbox{cellone/bulbC}}
\put(0, 240)
{\epsfxsize=63pt \epsfbox{celltwo/bulbC}}
\put(0, 160)
{\epsfxsize=63pt \epsfbox{cellthree/bulbC}}
\put(0, 80)
{\epsfxsize=63pt \epsfbox{cellfour/bulbC}}
\put(0, 0)
{\epsfxsize=63pt \epsfbox{cellfive/bulbC}}

\put(110, 400) {\footnotesize {Cortical}}
\put(110, 380) {\footnotesize {outputs}}
\put(100, 320)
{\epsfxsize=63pt \epsfbox{cellone/ctexC}}
\put(100, 240)
{\epsfxsize=63pt \epsfbox{celltwo/ctexC}}
\put(100, 160)
{\epsfxsize=63pt \epsfbox{cellthree/ctexC}}
\put(100, 80)
{\epsfxsize=63pt \epsfbox{cellfour/ctexC}}
\put(100, 0)
{\epsfxsize=63pt \epsfbox{cellfive/ctexC}}
\end{picture}
}
\put(200,360){\bf B}
\put(200,0)
{
\setlength{\unitlength}{0.7pt}
\begin{picture}(200, 400)
\put(10, 400) {\footnotesize {Bulbar}}
\put(10, 385) {\footnotesize {outputs}}
\put(0, 320)
{\epsfxsize=63pt \epsfbox{cellone/bulbB}}
\put(0, 240)
{\epsfxsize=63pt \epsfbox{celltwo/bulbB}}
\put(0, 160)
{\epsfxsize=63pt \epsfbox{cellthree/bulbB}}
\put(0, 80)
{\epsfxsize=63pt \epsfbox{cellfour/bulbB}}
\put(0, 0)
{\epsfxsize=63pt \epsfbox{cellfive/bulbB}}

\put(110, 400) {\footnotesize {Cortical}}
\put(110, 380) {\footnotesize {outputs}}
\put(100, 320)
{\epsfxsize=63pt \epsfbox{cellone/ctexB}}
\put(100, 240)
{\epsfxsize=63pt \epsfbox{celltwo/ctexB}}
\put(100, 160)
{\epsfxsize=63pt \epsfbox{cellthree/ctexB}}
\put(100, 80)
{\epsfxsize=63pt \epsfbox{cellfour/ctexB}}
\put(100, 0)
{\epsfxsize=63pt \epsfbox{cellfive/ctexB}}
\end{picture}
}
\put(0,360){\bf A}
\put(0,0)
{
\setlength{\unitlength}{0.7pt}
\begin{picture}(200, 400)
\put(10, 400) {\footnotesize {Bulbar}}
\put(10, 385) {\footnotesize {outputs}}
\put(0, 320)
{\epsfxsize=63pt \epsfbox{cellone/bulbA}}
\put(0, 240)
{\epsfxsize=63pt \epsfbox{celltwo/bulbA}}
\put(0, 160)
{\epsfxsize=63pt \epsfbox{cellthree/bulbA}}
\put(0, 80)
{\epsfxsize=63pt \epsfbox{cellfour/bulbA}}
\put(0, 0)
{\epsfxsize=63pt \epsfbox{cellfive/bulbA}}

\put(110, 400) {\footnotesize {Cortical}}
\put(110, 380) {\footnotesize {outputs}}
\put(100, 320)
{\epsfxsize=63pt \epsfbox{cellone/ctexA}}
\put(100, 240)
{\epsfxsize=63pt \epsfbox{celltwo/ctexA}}
\put(100, 160)
{\epsfxsize=63pt \epsfbox{cellthree/ctexA}}
\put(100, 80)
{\epsfxsize=63pt \epsfbox{cellfour/ctexA}}
\put(100, 0)
{\epsfxsize=63pt \epsfbox{cellfive/ctexA}}
\end{picture}
}
\end{picture}
\end{center}
\caption{ \label{fig:3odors} 
\footnotesize{{\bf A, B, C}: bulbar and cortical oscillation 
patterns for odors A, B (stored in the associative memory in the cortex) 
and C (not stored).
In each pattern, we plot temporal traces of outputs from 5 of the 50 mitral or 
cortical excitatory units during one sniff cycle lasting 370 milliseconds, 
roughly the first half of which is inhalation. Note the modulating of 
oscillation by the sniff cycle, and the different oscillation amplitudes for 
different units.  Oscillation phases also differ between units, though they 
are not apparent in the figure. The same format is used to display bulbar 
and cortical responnses in the following figures.  Cortex-to-bulb feedback 
is turned off for the results shown in this figure.  Note that the cortex 
responds little to odor C, since the input does not match any of the stored 
oscillation patterns.
} } 
\end{figure}

Fig.\ \ref{fig:adaptA} demonstrates odor adaptation to odor A.
The response amplitudes decay quicky in successive sniffs, although
the oscillation patterns do not change appreciably before the 
amplitudes decay to zero. The way this comes about is that the 
feedback signal generated by A, when relayed by the granule cells to
the mitral ones, creates an effective extra input signal $\bar {\rm A}$
(anti-A), and by the third sniff ${\rm A} +\bar {\rm A} \approx 0$.

To quantify the similarity between oscillation patterns, we extract 
an $N$=50 dimensional complex vector $\bf O$ from the temporal Fourier transform 
of the activity of the cortical excitatory units during the sniff cycle, 
with the component $O_i$ specifying the amplitude and phase of the 
oscillations in excitatory unit $i$.
We can measure the similarity between $\bf O$ and $\bf O'$ by the normalized 
overlap 
$S_{\rm OO'} = |\langle {\bf O}|{\bf O'}\rangle/(|{\bf O}|\cdot |{\bf O'}|)$, 
which is 1 for
${\bf O} \propto {\bf O'}$ and near zero (${\rm O}(1/\sqrt{N})$)
for two unrelated
patterns.  Calling the pattern vectors ${\bf A}^0$, 
${\bf A}^1$, ${\bf A}^2$, and ${\bf A}^3$ for 
cortical response to odor A without adaptation and during the first, 
second, and third sniff cycles of the adaptation respectively, we find
$S_{{\rm A}^0{\rm A}^1} = 0.9997$, $S_{{\rm A}^0{\rm A}^2} = 0.992$, 
and $S_{{\rm A}^0{\rm A}^3} = 0.74$, 
with response amplitudes $|{\bf A}^1|/|{\bf A}^0| = 0.97$, 
$|{\bf A}^2|/|{\bf A}^0| = 0.3$, $|{\bf A}^3|/|{\bf A}^0| = 0.08$.  Thus, 
the strength of the response is already significantly weakened after one 
sniff, but its cortical pattern of variation remains undistorted through 
several sniffs.

The way this adaptation varies in successive sniffs depends on both the
time constants in the feedback circuitry (as discussed above) and the 
strength of the filtered signal fed back to the bulb.   In the simulations
shown here, the latter was strong enough that even after one sniff, a large
fraction of the input signal is cancelled by the feedback, and after two
sniffs the cancellation was nearly complete.  A smaller feedback strength and
a correspondingly longer time constant of the feedback circuitry would
make it take longer for the adaptation to set in.  Similarly, the time it
takes for the adaptation, once established, to wear off is set by the same
time constants (for the values used here, around 3 s or 12 sniff cycles).

\begin{figure}[hthththth!]
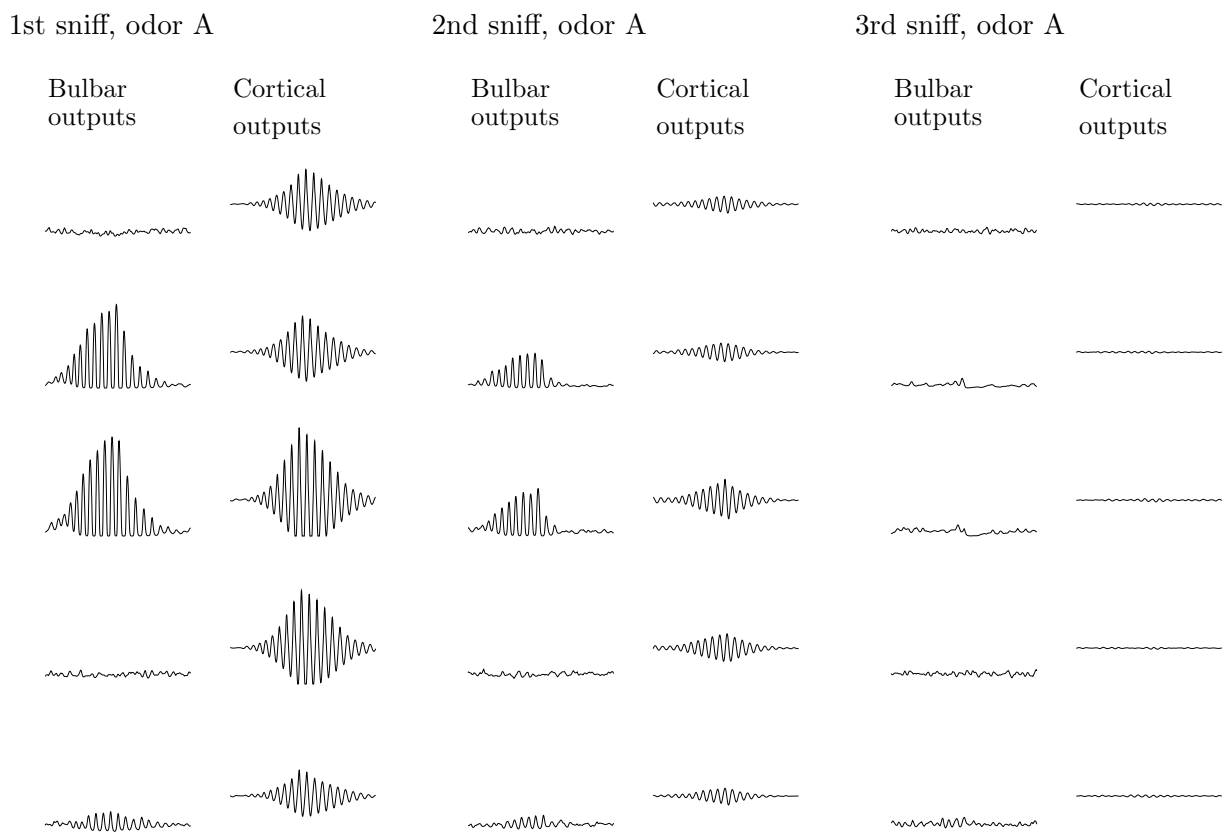

\begin{center}
\setlength{\unitlength}{0.8pt}
\begin{picture}(600, 380)
\put(400,380) {\small 3rd sniff, odor A}
\put(400,0)
{
\setlength{\unitlength}{0.7pt}
\begin{picture}(200, 400)
\put(10, 400) {\footnotesize {Bulbar}}
\put(10, 385) {\footnotesize {outputs}}
\put(0, 320)
{\epsfxsize=63pt \epsfbox{cellone/bulbAfb3}}
\put(0, 240)
{\epsfxsize=63pt \epsfbox{celltwo/bulbAfb3}}
\put(0, 160)
{\epsfxsize=63pt \epsfbox{cellthree/bulbAfb3}}
\put(0, 80)
{\epsfxsize=63pt \epsfbox{cellfour/bulbAfb3}}
\put(0, 0)
{\epsfxsize=63pt \epsfbox{cellfive/bulbAfb3}}
\put(110, 400) {\footnotesize {Cortical}}
\put(110, 380) {\footnotesize {outputs}}
\put(100, 320)
{\epsfxsize=63pt \epsfbox{cellone/ctexAfb3}}
\put(100, 240)
{\epsfxsize=63pt \epsfbox{celltwo/ctexAfb3}}
\put(100, 160)
{\epsfxsize=63pt \epsfbox{cellthree/ctexAfb3}}
\put(100, 80)
{\epsfxsize=63pt \epsfbox{cellfour/ctexAfb3}}
\put(100, 0)
{\epsfxsize=63pt \epsfbox{cellfive/ctexAfb3}}
\end{picture}
}
\put(200,380){\small 2nd sniff,  odor A}
\put(200,0)
{
\setlength{\unitlength}{0.7pt}
\begin{picture}(200, 400)
\put(10, 400) {\footnotesize {Bulbar}}
\put(10, 385) {\footnotesize {outputs}}
\put(0, 320)
{\epsfxsize=63pt \epsfbox{cellone/bulbAfb2}}
\put(0, 240)
{\epsfxsize=63pt \epsfbox{celltwo/bulbAfb2}}
\put(0, 160)
{\epsfxsize=63pt \epsfbox{cellthree/bulbAfb2}}
\put(0, 80)
{\epsfxsize=63pt \epsfbox{cellfour/bulbAfb2}}
\put(0, 0)
{\epsfxsize=63pt \epsfbox{cellfive/bulbAfb2}}
\put(110, 400) {\footnotesize {Cortical}}
\put(110, 380) {\footnotesize {outputs}}
\put(100, 320)
{\epsfxsize=63pt \epsfbox{cellone/ctexAfb2}}
\put(100, 240)
{\epsfxsize=63pt \epsfbox{celltwo/ctexAfb2}}
\put(100, 160)
{\epsfxsize=63pt \epsfbox{cellthree/ctexAfb2}}
\put(100, 80)
{\epsfxsize=63pt \epsfbox{cellfour/ctexAfb2}}
\put(100, 0)
{\epsfxsize=63pt \epsfbox{cellfive/ctexAfb2}}
\end{picture}
}
\put(0,380){\small 1st sniff, odor A}
\put(0,0)
{
\setlength{\unitlength}{0.7pt}
\begin{picture}(200, 400)
\put(10, 400) {\footnotesize {Bulbar}}
\put(10, 385) {\footnotesize {outputs}}
\put(0, 320)
{\epsfxsize=63pt \epsfbox{cellone/bulbAfb}}
\put(0, 240)
{\epsfxsize=63pt \epsfbox{celltwo/bulbAfb}}
\put(0, 160)
{\epsfxsize=63pt \epsfbox{cellthree/bulbAfb}}
\put(0, 80)
{\epsfxsize=63pt \epsfbox{cellfour/bulbAfb}}
\put(0, 0)
{\epsfxsize=63pt \epsfbox{cellfive/bulbAfb}}

\put(110, 400) {\footnotesize {Cortical}}
\put(110, 380) {\footnotesize {outputs}}
\put(100, 320)
{\epsfxsize=63pt \epsfbox{cellone/ctexAfb}}
\put(100, 240)
{\epsfxsize=63pt \epsfbox{celltwo/ctexAfb}}
\put(100, 160)
{\epsfxsize=63pt \epsfbox{cellthree/ctexAfb}}
\put(100, 80)
{\epsfxsize=63pt \epsfbox{cellfour/ctexAfb}}
\put(100, 0)
{\epsfxsize=63pt \epsfbox{cellfive/ctexAfb}}

\end{picture}
}

\end{picture}
\end{center}
\caption{ \label{fig:adaptA}
\footnotesize{Demonstrating the adaptation to odor A, with the feedback
from cortex to bulb active.  Plotted are the responses to odor A alone
during three successive sniffs.
Note that the response magnitudes decay in successive sniffs, but the 
response pattern, in particular, the relative amplitude pattern, stays
roughly the same from the first to second sniff before responses
disappear at the third sniff.
}
}
\end{figure}

Fig.\ \ref{fig:crossadapt}a demonstrates the segmentation capability of
the system.  The response ${\bf B}^{\rm seg}$ to the odor mixture A+B at the 
third sniff after the first 2 sniffs of odor A is quite similar to that, 
${\bf B}^0$, to odor B alone:   $S_{{\rm B}^{\rm seg}{\rm B}^0} = 0.993$, and
$|{\bf B}^{\rm seg}|/|{\bf B}^0| =0.91$.   Thus, although A is still present, 
so is the anti-A, so the net signal to the mitral units is 
${\rm A} + \bar {\rm A}+{\rm B} \approx {\rm B}$.  This demonstrates 
odor-specific adaptation in the model.  The system responds with the 
activity pattern characterizing the new odor, essentially undistorted by 
the existing odors in the environment, thus effectively achieving odor
segmentation.  Odor B can be segmented as long as it enters after the adaptation
to A is established, in this model at the 3rd or any subsequent sniffs.

\begin{figure}[hthththt!]
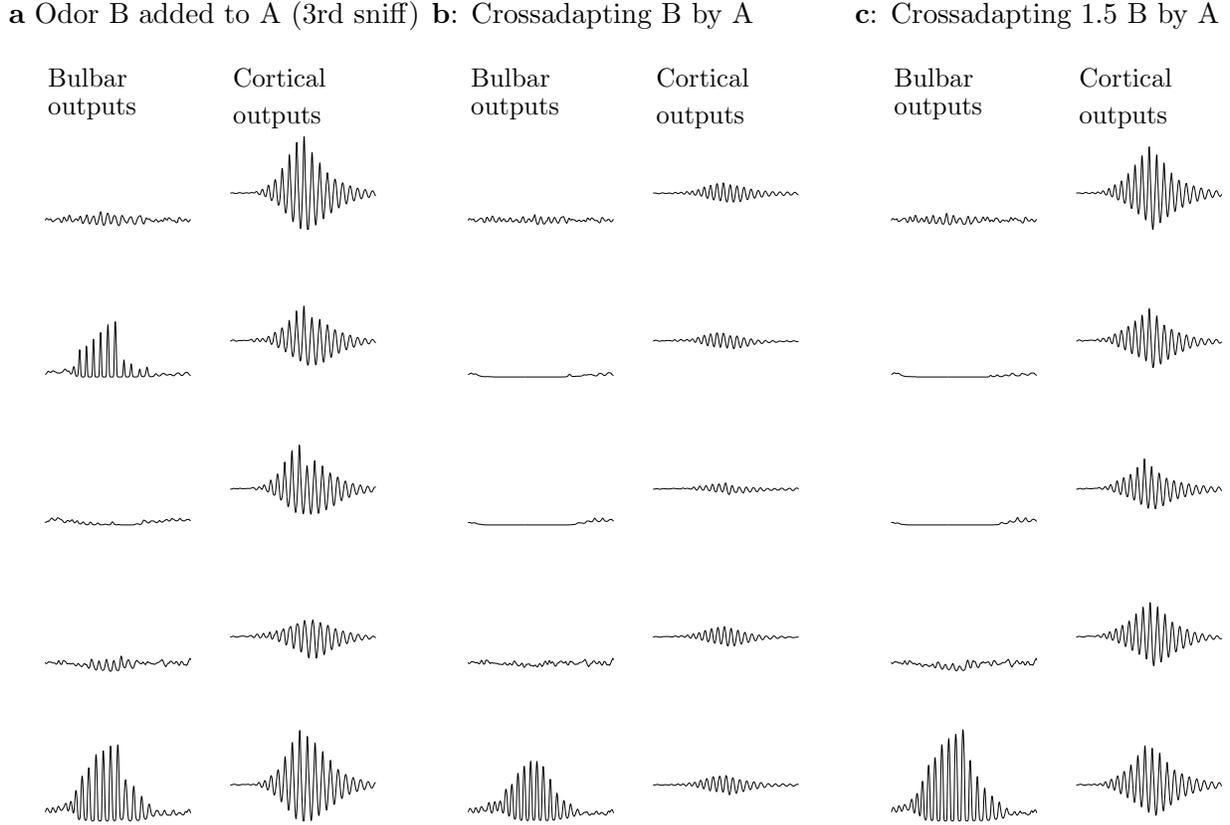

\begin{center}
\setlength{\unitlength}{0.8pt}
\begin{picture}(600, 380)(0, 0)
\put(400,380) {\small {\bf c}: Crossadapting 1.5 B by A}
\put(400,0)
{
\setlength{\unitlength}{0.7pt}
\begin{picture}(200, 400)
\put(10, 400) {\footnotesize {Bulbar}}
\put(10, 385) {\footnotesize {outputs}}
\put(0, 320)
{\epsfxsize=63pt \epsfbox{cellone/bulbA3moreB1subfb}}
\put(0, 240)
{\epsfxsize=63pt \epsfbox{celltwo/bulbA3moreB1subfb}}
\put(0, 160)
{\epsfxsize=63pt \epsfbox{cellthree/bulbA3moreB1subfb}}
\put(0, 80)
{\epsfxsize=63pt \epsfbox{cellfour/bulbA3moreB1subfb}}
\put(0, 0)
{\epsfxsize=63pt \epsfbox{cellfive/bulbA3moreB1subfb}}

\put(110, 400) {\footnotesize {Cortical}}
\put(110, 380) {\footnotesize {outputs}}
\put(100, 320)
{\epsfxsize=63pt \epsfbox{cellone/ctexA3moreB1subfb}}
\put(100, 240)
{\epsfxsize=63pt \epsfbox{celltwo/ctexA3moreB1subfb}}
\put(100, 160)
{\epsfxsize=63pt \epsfbox{cellthree/ctexA3moreB1subfb}}
\put(100, 80)
{\epsfxsize=63pt \epsfbox{cellfour/ctexA3moreB1subfb}}
\put(100, 0)
{\epsfxsize=63pt \epsfbox{cellfive/ctexA3moreB1subfb}}
\end{picture}
}
\put(200,380) {\small {\bf b}: Crossadapting B by A} 
\put(200,0) 
{ 
\setlength{\unitlength}{0.7pt} 
\begin{picture}(200, 400) 
\put(10, 400) {\footnotesize {Bulbar}} 
\put(10, 385) {\footnotesize {outputs}} 
\put(0, 320) 
{\epsfxsize=63pt \epsfbox{cellone/bulbA3B1subfb}}
\put(0, 240)
{\epsfxsize=63pt \epsfbox{celltwo/bulbA3B1subfb}}
\put(0, 160)
{\epsfxsize=63pt \epsfbox{cellthree/bulbA3B1subfb}}
\put(0, 80)
{\epsfxsize=63pt \epsfbox{cellfour/bulbA3B1subfb}}
\put(0, 0)
{\epsfxsize=63pt \epsfbox{cellfive/bulbA3B1subfb}}

\put(110, 400) {\footnotesize {Cortical}}
\put(110, 380) {\footnotesize {outputs}}
\put(100, 320)
{\epsfxsize=63pt \epsfbox{cellone/ctexA3B1subfb}}
\put(100, 240)
{\epsfxsize=63pt \epsfbox{celltwo/ctexA3B1subfb}}
\put(100, 160)
{\epsfxsize=63pt \epsfbox{cellthree/ctexA3B1subfb}}
\put(100, 80)
{\epsfxsize=63pt \epsfbox{cellfour/ctexA3B1subfb}}
\put(100, 0)
{\epsfxsize=63pt \epsfbox{cellfive/ctexA3B1subfb}}
\end{picture}
}

\put(0,380){\small {\bf a} Odor B added to A (3rd sniff)}
{
\setlength{\unitlength}{0.7pt}
\begin{picture}(200, 400)
\put(10, 400) {\footnotesize {Bulbar}}
\put(10, 385) {\footnotesize {outputs}}
\put(0, 320)
{\epsfxsize=63pt \epsfbox{cellone/bulbA3B1fb}}
\put(0, 240)
{\epsfxsize=63pt \epsfbox{celltwo/bulbA3B1fb}}
\put(0, 160)
{\epsfxsize=63pt \epsfbox{cellthree/bulbA3B1fb}}
\put(0, 80)
{\epsfxsize=63pt \epsfbox{cellfour/bulbA3B1fb}}
\put(0, 0)
{\epsfxsize=63pt \epsfbox{cellfive/bulbA3B1fb}}

\put(110, 400) {\footnotesize {Cortical}}
\put(110, 380) {\footnotesize {outputs}}
\put(100, 320)
{\epsfxsize=63pt \epsfbox{cellone/ctexA3B1fb}}
\put(100, 240)
{\epsfxsize=63pt \epsfbox{celltwo/ctexA3B1fb}}
\put(100, 160)
{\epsfxsize=63pt \epsfbox{cellthree/ctexA3B1fb}}
\put(100, 80)
{\epsfxsize=63pt \epsfbox{cellfour/ctexA3B1fb}}
\put(100, 0)
{\epsfxsize=63pt \epsfbox{cellfive/ctexA3B1fb}}
\end{picture}

}
\end{picture}
\end{center}
\caption{ \label{fig:crossadapt} \footnotesize{ a: Segmenting odors A and B.
After two sniffs of A as in Fig.\ \ref{fig:adaptA}, odor B is added, so
the net input is A+B.  The response is almost the same as that to B alone
(Fig.\ \ref{fig:3odors}, middle).   
b: Cross-adaption: response to odor B after odor A was present in two 
previous sniffs and then withdrawn.  The response is weak and distorted.
c: Same as b, except that an odor B 1.5 times as strong is used.  This
strength is sufficient to evoke a stronger, less distorted response.
}
}
\end{figure}

However, if odor A is suddenly withdrawn at the start of the 3rd sniff,
when odor B is introduced, the system response to odor B is weakened and 
distorted (this is particularly noticable in the bulbar responses).  The 
reason for this is that the effective total input is now 
${\rm B} + \bar{\rm A} \approx {\rm B} - {\rm A}$,
which is not at all like $\rm B$ (Fig.\ \ref{fig:crossadapt}, b and c).
This corresponds to the psychophysically observed cross-adaptation ---
after sniffing one odor, another odor at next sniff smells
less strong than it normally would and  may even smell different
\cite{Moncrieff}. In the normal olfactory environment, however, such
sudden and near complete withdrawal of an odor seldom happens. Let 
${\bf B}^{\rm cross}$ and ${\bf \tilde B}^{\rm cross}$ be the cortical response 
vectors to cross adapted odor B and odor 1.5B. Comparing with
the response to odor B alone, we find
$S_{{\rm B}^0 {\rm B}^{\rm cross}} = 0.94$, 
$|{\rm B}^{\rm cross}|/|{\rm B}^0| = 0.23$; 
$S_{{\rm B}^0 {\rm \tilde B}^{\rm cross}} = 0.97$, 
$|{\rm \tilde B}^{\rm cross}|/|{\rm B}^0| = 0.74$. 
We can understand these results in the following way.  The feedback input
$\bar{\rm A} \approx -{\rm A}$ acts to move the bulb operating point $\bar x_i$
to lower gain values (for units where $I_i^{\rm A}$ is strong), thereby
weakening the overall response.  For normal-strength B, most of the mitral 
units in the bulb do not respond much, so the cortical response is 
correspondingly weak and distorted relative to that to B in the absence of 
adaptation.  The stronger input 1.5B evokes a stronger bulb response, 
however, and the cortical response is stronger and better (but still 
imperfectly) correlated with the unadapted pattern. 

Since the olfactory bulb is nonlinear, the odor mixture A+B
does not induce a bulbar response equal to the sum
of the responses to A and B individually.  Consequently, the unadapted
cortical response to it (Fig.\ \ref{fig:adaptAB}, left panel) is weaker than 
that to A or B (the bulb response to the mixture is not embedded in the
cortical connections) and not strongly correlated with the responses to
the pure odors.  The situation is similar to that for any other 
unstored odor, such as C (Fig.\ \ref{fig:3odors}, panel C), to which there 
is almost no adaptation in the bulb because there is almost no cortical
signal to feed back.  The unadapted cortical response to A+B is stronger 
than that to C because the nonlinearity in the bulb here is not strong
enough to completely destroy correlations between its reponses to the
individual odors A and B and that to their mixture.  Nevertheless, 
the weakness of the cortical response reduces the feedback to the bulb
significantly, and the system does not adapt to the mixture as 
effectively as to individual odors, as shown in the middle and right panels 
of Fig.\ \ref{fig:adaptAB}.  We also note that because the feedback is weak,
the attenuation of the signals in both bulb and cortex, is also weaker than 
for pure stored odors (cf Fig.\ \ref{fig:adaptA}).  Thus, the cortical
response to the mixed odor, while initially weaker than that to pure 
stored ones, lasts longer.

\begin{figure}[hthththt!]
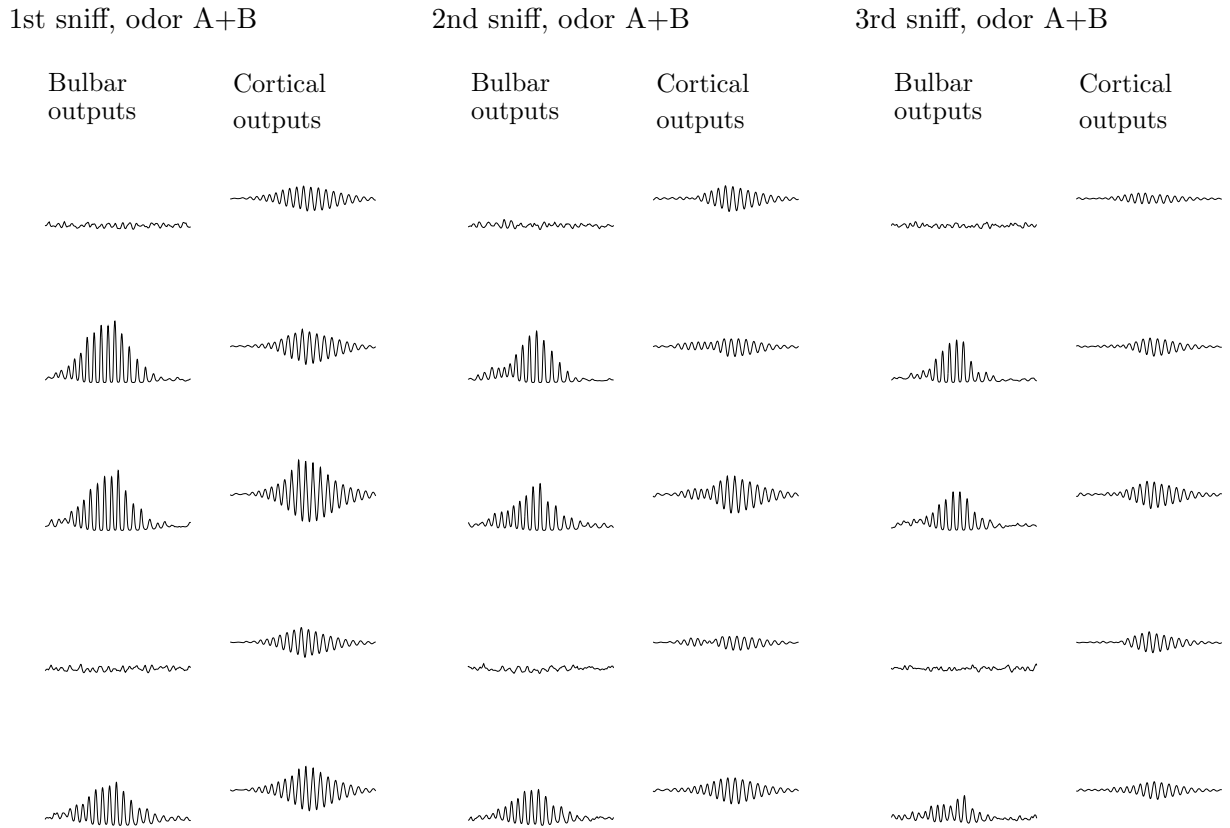

\begin{center}
\setlength{\unitlength}{0.8pt}
\begin{picture}(600, 380)
\put(400,380) {\small 3rd sniff, odor A+B}
\put(400,0)
{
\setlength{\unitlength}{0.7pt}
\begin{picture}(200, 400)
\put(10, 400) {\footnotesize {Bulbar}}
\put(10, 385) {\footnotesize {outputs}}
\put(0, 320)
{\epsfxsize=63pt \epsfbox{cellone/bulbABfb3}}
\put(0, 240)
{\epsfxsize=63pt \epsfbox{celltwo/bulbABfb3}}
\put(0, 160)
{\epsfxsize=63pt \epsfbox{cellthree/bulbABfb3}}
\put(0, 80)
{\epsfxsize=63pt \epsfbox{cellfour/bulbABfb3}}
\put(0, 0)
{\epsfxsize=63pt \epsfbox{cellfive/bulbABfb3}}

\put(110, 400) {\footnotesize {Cortical}}
\put(110, 380) {\footnotesize {outputs}}
\put(100, 320)
{\epsfxsize=63pt \epsfbox{cellone/ctexABfb3}}
\put(100, 240)
{\epsfxsize=63pt \epsfbox{celltwo/ctexABfb3}}
\put(100, 160)
{\epsfxsize=63pt \epsfbox{cellthree/ctexABfb3}}
\put(100, 80)
{\epsfxsize=63pt \epsfbox{cellfour/ctexABfb3}}
\put(100, 0)
{\epsfxsize=63pt \epsfbox{cellfive/ctexABfb3}}
\end{picture}
}
\put(200,380){\small 2nd sniff,  odor A+B}
\put(200,0)
{
\setlength{\unitlength}{0.7pt}
\begin{picture}(200, 400)
\put(10, 400) {\footnotesize {Bulbar}}
\put(10, 385) {\footnotesize {outputs}}
\put(0, 320)
{\epsfxsize=63pt \epsfbox{cellone/bulbABfb2}}
\put(0, 240)
{\epsfxsize=63pt \epsfbox{celltwo/bulbABfb2}}
\put(0, 160)
{\epsfxsize=63pt \epsfbox{cellthree/bulbABfb2}}
\put(0, 80)
{\epsfxsize=63pt \epsfbox{cellfour/bulbABfb2}}
\put(0, 0)
{\epsfxsize=63pt \epsfbox{cellfive/bulbABfb2}}

\put(110, 400) {\footnotesize {Cortical}}
\put(110, 380) {\footnotesize {outputs}}
\put(100, 320)
{\epsfxsize=63pt \epsfbox{cellone/ctexABfb2}}
\put(100, 240)
{\epsfxsize=63pt \epsfbox{celltwo/ctexABfb2}}
\put(100, 160)
{\epsfxsize=63pt \epsfbox{cellthree/ctexABfb2}}
\put(100, 80)
{\epsfxsize=63pt \epsfbox{cellfour/ctexABfb2}}
\put(100, 0)
{\epsfxsize=63pt \epsfbox{cellfive/ctexABfb2}}
\end{picture}
}
\put(0,380){\small 1st sniff, odor A+B}
\put(0,0)
{
\setlength{\unitlength}{0.7pt}
\begin{picture}(200, 400)
\put(10, 400) {\footnotesize {Bulbar}}
\put(10, 385) {\footnotesize {outputs}}
\put(0, 320)
{\epsfxsize=63pt \epsfbox{cellone/bulbABfb}}
\put(0, 240)
{\epsfxsize=63pt \epsfbox{celltwo/bulbABfb}}
\put(0, 160)
{\epsfxsize=63pt \epsfbox{cellthree/bulbABfb}}
\put(0, 80)
{\epsfxsize=63pt \epsfbox{cellfour/bulbABfb}}
\put(0, 0)
{\epsfxsize=63pt \epsfbox{cellfive/bulbABfb}}

\put(110, 400) {\footnotesize {Cortical}}
\put(110, 380) {\footnotesize {outputs}}
\put(100, 320)
{\epsfxsize=63pt \epsfbox{cellone/ctexABfb}}
\put(100, 240)
{\epsfxsize=63pt \epsfbox{celltwo/ctexABfb}}
\put(100, 160)
{\epsfxsize=63pt \epsfbox{cellthree/ctexABfb}}
\put(100, 80)
{\epsfxsize=63pt \epsfbox{cellfour/ctexABfb}}
\put(100, 0)
{\epsfxsize=63pt \epsfbox{cellfive/ctexABfb}}

\end{picture}
}

\end{picture}
\end{center}
\caption{ \label{fig:adaptAB} \footnotesize{This figure illustrates how
adaptation in the model is not effective for the mixture odor (A+B)/2.
Responses to this odor are shown for 3 successive sniff cycles. The 
cortical response, although initially weaker than that to pure A or B
(Fig.\ \ref{fig:3odors}) is still appreciable
at the 3rd sniff (compare with adaptation to odor A in Fig.\ 
\ref {fig:adaptA}).}}
\end{figure}

\section{Discussion}

We have presented a computational model for an olfactory system 
that can detect, recognize and segment odors.  Detection is performed 
in the bulb, which encodes odors in oscillatory activity patterns.  
Recognition is carried out by the cortex using a resonant associative 
memory mechanism.  Finally, segmentation is implemented by a slowly-varying 
feedback signal which acts to cancel the specific input that 
evoked the resonant cortical response.   

The model is constrained by a few basic anatomical and physiological 
facts:  Odors evoke oscillatory activity in populations of excitatory 
and inhibitory neurons in both bulb and cortex, these two structures 
are coupled by both feedforward and feedback connections, reducing the
cortical feedback enhances the bulbar responses,  and the system
exhibits odor-specific adaptation.  Within these constraints, we have
tried to build a minimal model.  We have taken the bulb module from 
earlier work by one of us \cite{LH,Li90} and augmented it with a model
of the pyriform cortex and with feedforward and feedback connections 
between it and the bulb.  We have ignored many further known details 
of real olfactory systems that do not bear directly on the fundamental
property of stimulus-specific adaptation, and when we have had to go 
beyond current knowledge (as in constructing the feedback signal) 
we have done so in a purely phenomenological way, avoiding hypothesizing 
specific details unrelated to the function of the system.  From the
analysis of the model and the simulations we can see how the basic 
computations necessary for olfactory segmentation might be carried out
by the neural networks of the bulb and cortex. 

But do real olfactory systems actually function in this way?  This can
be tested at the level of both the assumptions we put into the model
and the properties we find for it.  First of all, we have assumed that
the feedback from cortex to bulb is slowly varying (i.e. that firing 
rates for the feedback fibers vary on the timescale of the sniff cycle,
but not of the oscillations found in both the bulb and cortex).  Furthermore,
we have assumed that this feedback is odor-specific.  While the existence
of some feedback is well-established, neither of these specific
hypotheses has been tested experimentally.  However they both could be.

Properties we find in the model, beyond the fact that it successfully 
implements segmentation, can also be tested.  These include the following:  

First, the fact that the feedback signal requires strong cortical activity
to drive it means that unfamiliar (unlearnt) odors will not be adapted to 
as strongly as familiar ones, so they will not be so easily segmented from 
subsequently presented ones.  As we saw in Fig.\ \ref{fig:adaptAB}, this 
expectation also applies to unfamiliar mixtures of familiar odors.  
Furthermore, as we also noted, we expect the weakening of the (initially 
weaker) responses with adaptation to be slower for such mixtures 
than for familiar odors.

Second, cross-adaptation, as illustrated in Fig.\ 
\ref{fig:crossadapt}, is a necessary consequence of the slow feedback:
The effective bulb input $\bar{\rm A} \approx -{\rm A}$, from the previous 
presence of the adapting stimulus, will be present for some time (depending
on the time constants of the feedback circuitry) whether odor A remains in 
the environment or not.  Thus the total input to the bulb with A still
present will be very different from that with A suddenly removed.  If
there is odor-specific adaptation of the kind necessary to perform 
segmentation when A remains in the environment (A cancelled by 
$\bar{\rm A}$), then a different response must occur when A is withdrawn.
Present evidence on cross-adaptation is rather limited, but psychophysical
and electrophysiological investigation of this phenomenon would be helpful
in pinning down quantitatively the time constants of the circuitry
involved in segmentation.

If odor-specific adaptation is not implemented using our cortical feedback 
mechanism, how else might it be done?  One possibility to consider is 
single-unit-level adaptation (or fatigue), which can be implemented in a 
network like ours by making the threshold for each unit dependent on its 
own recent activity.    In a model with the structure of ours (with bulb 
and cortical modules) but without feedback, such fatigue would have to be 
implemented in the bulb; otherwise the activity there would not exhibit
adaptation.  This presents a problem if the activity patterns of different
odors overlap significantly -- it is not evident that one can avoid changing 
the response to a new odor when some of the units active in the normal
response to it are to be fatigued.  Indeed, in investigations of simple
oscillatory associative memories with such adaptation \cite{Wangetal},
temporal segmentation has been found only for patterns with rather weak
mutual overlap.  This overlap will be weak for sparse patterns, but it
is not clear how sparse real evoked bulb and cortical activity patterns
are, when looked at at the level of resolution of the units in our model.  

This problem is not present for the mechanism we propose, in which bulb
units themselves are not fatigued.  Rather, the mechanism cancels the 
input to bulb units in exactly the degree that they receive input 
from the adapting odor.  It is as if every receptor activated by an odor
became adapted by an amount exactly equal to its initial response.  

In our model, the feedback connections to the inhibitory bulb units have 
to have just the right values to produce the necessary cancellation.  In
real olfactory systems, the strengths of the centrifugal synapses on 
granule cells are presumably determined by some learning mechanism, and
for our model to apply it is necessary that this mechanism find the right
values for them.   As we know nothing about this mechanism, here 
in our model we just assumed the necessary form.  This form has a degree 
of plausibility because it is Hebbian, but very little is known yet about 
learning in these synapses.  Investigations could shed important light
on the validity of this key element of the model.

Another plausible mechanism, which could implement odor-specific adaptation
in the bulb in more or less the right manner, is adaptation of
receptor-bulb synapses in such a way that the inputs to bulb capture mainly the
transient but not static odor inputs.  
This would reduce the input signal for the 
adapting odor directly, at just the right places, and so does not suffer 
from the problems that single-unit fatigue in the bulb does.  However,
there is a simple difference between the predictions of such a model and 
ours, since in ours the cortex, functioning as an associative memory, only
sends its feedback to the bulb (or only sends it at full strength) for learnt 
odors.  The receptor-mitral synaptic adaptation model would exhibit the
same degree of odor-specific adaptation for all odors, learnt or not.  Of 
course, both mechanisms could be present, and the difference could be
large or small according to the relative size of the two contributions.

The fact that we have employed both excitatory-to-excitatory ($\sf J$)
and excitatory-to-inhibitory ($\sf \tilde W$) cortical connections enhances the 
associative memory function by permitting oscillation patterns which
differ in phase as well as amplitude.  This is of no help for 
selective adaptation in the model as described here, since phase information 
is lost in the generation of the feedback signal, but this information
could be retained using more elaborate mechanisms, as mentioned 
in Sect.\ 2.3.  

It is not clear whether real olfactory systems code odors in the phases
of their oscillation patterns.  However, in any case, a restricted version
of our cortex, without $\sf \tilde W$, could function with only 
amplitude-modulated patterns, similarly to the model of Wang et al 
\cite{Wangetal}.  The addition of intrinsic oscillatory properties for
individual units or, implicitly, the individual neurons in the populations
they represent, would not change the properties of such a network 
qualitatively.
 
The three tasks carried out by the system -- detection, recognition,
and segmentation -- are computationally linked.  For example, even
if an ambiguous or weak odor is ``recognized'' by the pyriform cortex
in the sense that a characteristic oscillatory response is evoked there,
that response may be too weak to suppress further bulbar response.
Then the system will continue to respond to the odor in the same way
as if it had not recognized it; that is, the odor-specific adaptation
necessary for segmentation can be seen as part of the recognition process.
  
While our units correspond to functional groups of neurons in
real olfactory systems,  our model is of higher resolution 
than that of Ambros-Ingerson {\em et al} \cite{AGL}.  While we emphasize 
the coding of odor information in distributed oscillation patterns, their 
model contains no explicit treatment of dynamics on the 40-hz timscale
or of the temporal segmentation problem.  They address instead a 
higher-level problem (hierarchical odor classification) with a 
higher-level model.  In such more complex situations, cortex-to-bulb
feedback could be a more general, active phenomenon than in the 
limited-scope problem we consider, but we do not address such issues
here.
 
Our network performs what might be called ``the simplest
cognitive computation".  It is natural to expect that evolution 
has employed elaborations on this structure in other sensory
systems and in central processing.  For example, hippocampal 
processing also employs oscillations, long-range intra-area associative
connections, and feedback \cite{RT,Hasselmorev}.  In another context, 
work by one of us \cite{LiNC} on visual processing suggests a function
for slow feedback to inhibitory neurons from higher areas in modulating the 
computations carried out in area V1.  Our hope is that studying and 
modeling the olfactory system in the way we have done here will lead to
insights into aspects of top-down/bottom-up interactions in other cognitive 
computations.

\section*{Appendix: Bulb-cortex coupling: implementation details}

\subsection*{Feedforward}

In the feedforward pathway from bulb to cortex, the mitral unit
outputs $g_x(x_i)$ are fed both directly to the excitatory cortical
units and in parallel, indirectly via feedforward inhibitory units.
The process, as indicated schematically in Fig. \ref{fig:olfactorysystem}, 
is described by the equations
\bea
L_i 	  &=& \sum_j C^{\rm b\rightarrow c}_{ij} g_x(x_j)   \label{eq:L} 	\\
\dot z_i    &=&  -\alpha_{\rm ff} z_i + L_i \label{eq:z}			\\ 
I^{\rm b}_i &=& L_i - \sigma g_z(z_i). 		    \label{eq:netin}
\eea
Here $L_i$ is the input signal to the cortical location $i$, 
${\sf C}^{\rm b\rightarrow c}$ is the connection matrix that transforms the 
mitral outputs to the cortical inputs, $z_i$ are the membrane potentials of
the feedforward inhibitory units, $g_z(.)$ is their activation function
and $\alpha_{\rm ff}^{-1}$ is their time constant.   $I^{\rm b}_i$ is then 
the total input signal to the $i$-th cortical excitatory unit in 
Eqn.\ (\ref{eq:ueqn}).  In general, this input contains both slowly-varying 
and rapidly-oscillating components.  The pathway via the inhibitory
feedforward units acts like a low-pass filter.  Thus, the net effect is
that the rapidly-varying or high frequency components, which contain the odor information, 
are transmitted to the cortex.  

In the simulations reported in Sect.\ 3, we took $g_z(.)$ to have two 
regions of different gain values, with a smaller gain at smaller input.
We designed $\sigma$ and the parameters of $g_z(.)$ so that the net slow 
component of $I^{\rm b}_i$  pushed the cortical operation points $\bar u_i$ 
and $\bar v_i$ to stable values close to, but below, their high gain 
region.   Thus the cortex had a stable operating point, enabling it to
carry out its associative memory function more cleanly that without this
engineering refinement.  

We make no claims about biological realism for the details of the feedfoward
mechanism.  However, some kind of effective high-pass filter is essential to
the robust functioning of the model.  Further experimental investigation of 
the dynamical properties of the feedfoward pathway would be important for 
understanding how it actually works.

\begin{figure}[ththththh!]
\begin{center}
\begin{picture}(200, 400)
\put(-50, 0)
{\epsfxsize=300pt \epsfbox{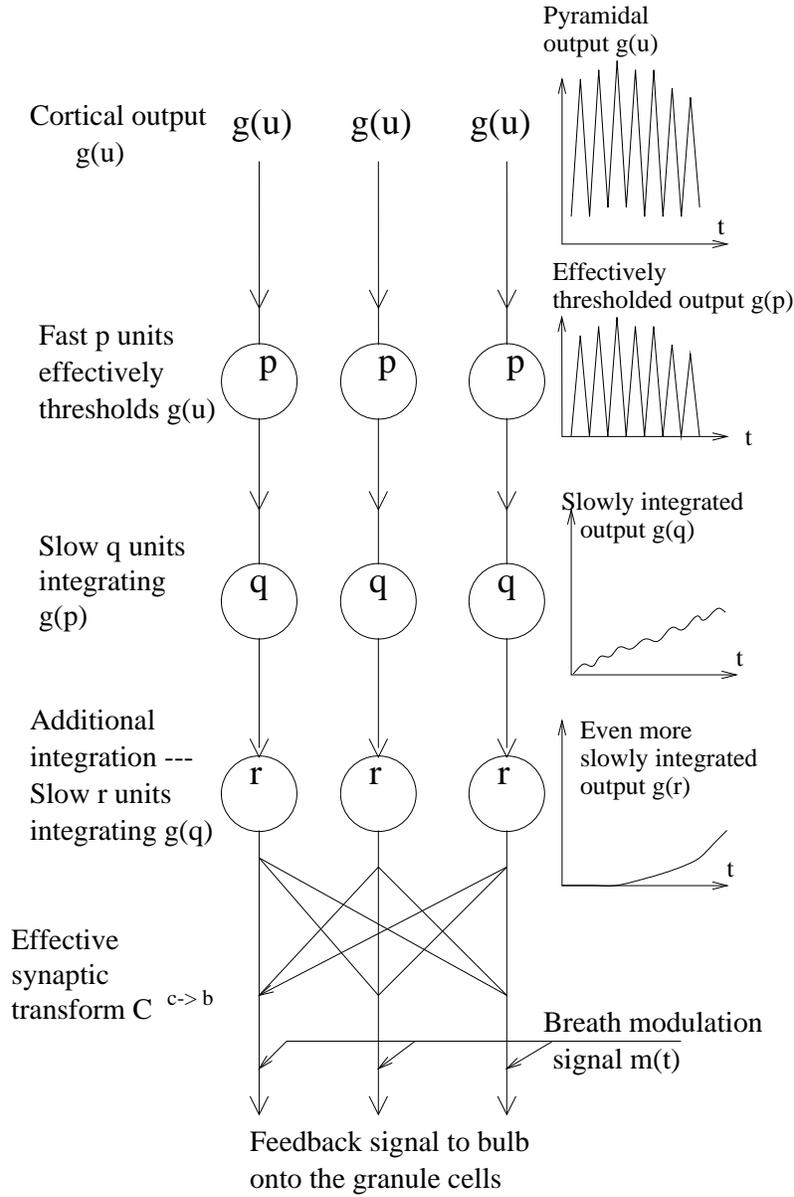}}
\end{picture}
\end{center}
\caption{ \label{fig:feedbackroute}
\footnotesize{Details of the feedback route. Only the oscillatory
components of the cortical outputs $g_u(u)$ contain the odor information. 
This component is extracted by half-wave rectification by the $p$ units.
The oscillatory $g_p(p)$ is converted to slowly varying signals by two
successive slow temporal integrating units $q$ and $r$.
The resulting signal $r(t)$ is fed through the matrix  
${\sf C}^{\rm c\rightarrow b}$ and modulated with the breathing cycle by a 
signal $m(t)$ to produce the odor-specific feedback signal to the bulbar 
granule units. The temporal characteristics of signals from different units 
are depicted schematically on the right.
}
}
\end{figure}
 
\subsection*{Feedback}

To generate the half-wave rectified, low-passed feedback signal to the bulb 
from the cortical excitatory unit outputs, we use three successive groups of 
units followed by a synaptic matrix, as shown in Fig.\ \ref{fig:feedbackroute}:
\begin{eqnarray}
        \dot p_i &=& -\alpha_{\rm fast} p_i + g_u(u_i),    \;\;\;
        \dot q_i = -\alpha_{\rm slow} q_i + g_p(p_i),    \;\;\;
        \dot r_i = -\alpha '_{\rm slow} r_i + q_i ,  \label{eq:pqr}    \\
        I^{\rm c}_i  &=&
m(t) \sum_j C^{\rm c\rightarrow b}_{ij} g_r(r_j), 	\label{eq:Ic}
\end{eqnarray}
where $m(t)$ is a modulating signal that synchronizes with breathing,
increasing during inhalation and decreasing during exhalation.

With a short time constant $1/\alpha_{\rm fast}$ and a strong nonlinear
$g_p$, the $p_i$ unit has a output $g_p(p_i)$ which is effectively 
 $g_u(u_i)$ thresholded above the average signal level.
This ``rectified''  output is then transformed by the 
two subsequent units $q_i$ and $r_i$, both with long time constants
$1/\alpha_{\rm slow}$ and $1/\alpha '_{\rm slow}$,  into a slowly-varying
signal, which is modulated by a function $m(t)$ (representing the breathing 
rhythm of the animal) and fed back via the connections 
${\sf C}^{\rm c \rightarrow b}$ to produce the centrifugal input 
${\bf I}^{\rm c}$ to the granule units in the bulb.  

It is not necessary to use two low-pass filter operations; the model 
works qualitatively the same with just one.  However, adding the second
one delays the feedback signal somewhat, giving the oscillations time
to establish themselves before the feedback begins to act.

In a more complete model, the large time constants $1/\alpha_{\rm slow}$ 
and $1/\alpha '_{\rm slow}$ could emerge as a dynamic network property of 
secondary olfactory areas. Similarly, the modulating signal $m(t)$ could 
arise from additional signals from other parts of the brain.


\begin{thebibliography}{99}

\bibitem{Laing} D. G. Laing
Perception of odor mixtures.
in {\it Handbook of olfaction and gustation} Ed. R. L. Doty, Marcel Dekker,
Inc. 1995. p 283-298.

\bibitem{HornUsher} D.\ Horn and M .\ Usher,  Parallel activation of memories 
in an oscillatory neural network, {\em Neural Comp} {\bf 3} 31-43 (1991)

\bibitem{HSU} D.\ Horn, D.\ Sagi and M.\ Usher, Segmentation, binding and 
illusory conjunctions, {\em Neural Comp} {\bf 3} 510-525 (1991)

\bibitem {Shepherd90}
G. M. Shepherd,  
Computational structure of the olfactory system,
in {\it Olfaction --- A Model System for
Computational Neuroscience } Ed J L Davis and H Eichenbaum, p 225-250,
MIT Press (1990)

\bibitem {Shepherd7990}
G M Shepherd  
{\it The Synaptic Organization of the Brain},
3rd edition, Oxford University Press (1990)

\bibitem {Freeman78}
W J Freeman, 
Spatial properties of an EEG event in the
olfactory bulb and cortex, {\it Electroencephalogr Clin Neurophysiol}
{\bf 44}, 586-605 (1978)

\bibitem {FS}
W J Freeman and W Schneider, Changes in spatial patterns of rabbit
olfactory EEG with conditioning to odors,
{\it Psychophysiology} {\bf 19}, 44-56 (1982)


\bibitem{FreemanGrajski}
W J Freeman and K A Grajski, 
Relation of olfactory EEG to behavior: 
factor analysis,
{\it Behav Neurosci} {\bf 101}, 766-77 (1987)

\bibitem {Adrian}
E D Adrian,
Sensory discrimination with some recent evidence from
the olfactory organ, 
{\it Br Med Bull} {\bf 6}, 330-331 (1950).

\bibitem{FreemanSkarda}
W J Freeman and C A Skarda, 
Spatial EEG patterns, non-linear dynamics and perceptio: 
the Neo-Sherrington view, {\it Brain Res Rev}
{\bf 10}, 147-175 (1985).

\bibitem{GelperinTank}
A Gelperin and D W Tank,
Odour-modulated collective network oscillations of olfactory 
interneurons in a terrestrial mollusc,
{\it Nature} {\bf 345}, 437-40 (1990)


\bibitem{Delaneyetal}
K R Delaney, A Gelperin, M S Fee, J A  Flores, R Gervais, 
D W Tank and D Kleinfeld,
Waves and stimulus-modulated dynamics in an oscillating 
                       olfactory network,
{\it Proc Natl Acad Sci USA} {\bf 91}, 669-73 (1994)  
 
\bibitem{Bressler}
S L Bressler, Changes in electrical activity of rabbit olfactory 
bulb and cortex to conditioned odor stimulation, {\em Behav Neurosci}
{\bf 102}, 740-747 (1988)

\bibitem {Moncrieff}
R W Moncrieff, {\it The Chemical Senses}, 3rd edition,
CRC Press (1967)

\bibitem {Maetal}
M Ma, T Leinders-Zufall, G M Sheperd, and F Zufall, 
Two forms of odor adaptation in single olfactory receptor neurons,
{\it Soc Neorosci Abstr} {\bf 23}, 741 (1997)

\bibitem {Haberly85}
L B Haberly, Neuronal circuitry in olfactory cortex:
anatomy and functional implications,
{\it Chem Senses} {\bf 10}, 219-238 (1985)

\bibitem{WB92}
M A Wilson and J D Bower, 
Cortical oscillations and temporal interactions in a 
computer simulation of piriform cortex,
{\it J Neurophysiol} {\bf 67}, 981-995 (1992)


\bibitem{AGL}
J Ambros-Ingerson, R Granger and G Lynch, Simulation of Paleocortex
Performs Hierarchical Clustering, {\em Science} {\bf 247}, 1344-1348 (1990)

\bibitem {Hasselmo} 
M. E. Hasselmo, 
Acetylcholine and learning in a cortical associative memory,
{\it Neural Computation} {\bf 5}, 32-44 (1993)


\bibitem{LilWu} H Liljenstr\"om and X-B Wu, 
Noise-enhanced performance in a cortical associative memory model,
{\em Int J Neural Systems} {\bf 6}, 19-29 (1995)

\bibitem{Lilj} H Liljenstr\"om,
Autonomous learning with complex dynamics,
{\em Int J Intelligent Systems} {\bf 10}, 119-153 (1995)

\bibitem{GraySkinner}
C M Gray and  J E Skinner, 
Centrifugal regulation of neuronal activity in the
olfactory bulb of the waking rabbit as revealed by
reversible cryogenic blockade,
{\em Exp Brain Res} {\bf 69}, 378-386 (1988)
 
\bibitem{WilsonCowan}
H R Wilson and J D Cowan, {\em Biophys J} {\bf 12}, 1-24 (1972)

\bibitem{LH}
Z Li and J Hopfield, 
A model of the olfactory bulb and its oscillatory processing,
{\it Biol Cybern} {\bf 61}, 379-392 (1989)

\bibitem{Li90}
Z Li, 
A model of odor adaptation and sensitivity enhancement in the
olfactory bulb,
{\it Biol Cybern} {\bf 62}, 349-361 (1990)

\bibitem{Hopfield}  J Hopfield, {\em Proc Nat Acad Sci USA} {\bf 79} 
2554-2558 (1982), {\bf 81} 3088-3092 (1984)

\bibitem{NR} W H Press, B P Flannery, S A Teukolsky, 
and W T Vetterling,
{\em Numerical Recipes in C}, Cambridge University Press (1988)



\bibitem{Wangetal}
D Wang, J Buhmann and C van der Malsburg, Pattern segmentation in associative
memory, {\em Neural Computation} {\bf 2}, 94-106 (1990)

\bibitem{RT}
E T Rolls and A Treves, {\em Neural Networks and Brain Function} Ch 6,
Oxford University Press (1998)

\bibitem{Hasselmorev}
M E Hasselmo, Neuromodulation and cortical function: modeling the
physiological basis of behavior, {\em Behav Brain Res} {\bf 67},
1-27 (1995)

\bibitem{LiNC}
Zhaoping Li, A neural model of contour integration in the primary
visual cortex, {\em Neural Comp} {\bf 10}, 903-940 (1998)

  
\end{thebibliography}
\end{document}